\documentclass[preprint]{aastex63}

\begin{document}



\title{Phase Shift of Planetary Waves and Wave--Jet Resonance on Tidally Locked Planets}

\author{Shuang Wang}
\affiliation{Dept. of Atmospheric and Oceanic Sciences, School of Physics, Peking University, Beijing 100871, China}

\author{Jun Yang}
\affiliation{Dept. of Atmospheric and Oceanic Sciences, School of Physics, Peking University, Beijing 100871, China}

\correspondingauthor{Jun Yang}
\email{junyang@pku.edu.cn}

\begin{abstract}
 Recent studies found that atmospheric superrotation (i.e., west-to-east winds over the equator) on tidally locked planets can modify the phase of planetary waves. But, a clear relationship between the superrotation and the magnitude of the phase shift was not examined. In this study, we re-investigate this problem using a two-dimensional (2D) linear shallow water model with a specified uniform zonal flow. We find that the degree of the phase shift is a monotonic but nonlinear function of the strength of the mean flow, and the phase shift has two limits of $-$$\pi$ and $+$$\pi$. The existence of these limits can be explained using the energy balance of the whole system. We further show that a resonance between the Rossby wave and the mean flow occurs when the speed of an eastward jet approaches to the westward phase speed of the Rossby wave, or a resonance between the Kelvin wave and the mean flow happens when the speed of a westward jet approaches to the eastward phase speed of the Kelvin wave. The resonance mechanism is the same as that found in the previous studies on Earth and hot Jupiters. Moreover, in the spin-up period of a 3D global atmospheric general circulation simulation for tidally locked rocky planet, we also find these two phenomena: phase shift and wave--jet resonance. This study improves the understanding of wave--mean flow interactions on tidally locked planets.
\end{abstract}



\section{Introduction}\label{session_introduction}

Phase curve observations of 1:1 tidally locked (or called synchronously rotating) hot jupiters showed that the hottest point of many (not all) hot jupiters is not at the substellar point and has an eastward shift \citep{Knutson_2007,Stevenson_2014,Heng_2015,Zhang_2018,Pierrehumbert_2018,Imamura_2020,Showman_2020}. For example, the phase curve of the hot Jupiter HD\,209458b demonstrated that its hottest spot is shifted obviously eastward of the substellar point by $40.9^{\circ}\pm6.0^{\circ}$\citep{Zellem_2014}. Theory studies and numerical simulations showed that the underlying reason is that atmospheric circulation on this type of planets is dominated by equatorial superrotation (i.e., west-to-east winds over the equator), which transports heat from the substellar region to the east \citep{Showman_2002,Showman_2011,Perez-Becker_2013}. When the radiative timescale is comparable to the advection or wave timescale, this superrotation is able to trigger an eastward shift of the hottest point.

The supperrotation is due to equator-ward momentum transports by coupled Rossby-Kelvin waves, which are ultimately excited from the uneven distribution of stellar energy between the permanent dayside and nightside \citep{Showman_2011}. In spatial pattern, a chevron-shape wave structure, northwest-to-southeast tilting in northern hemisphere and southwest-to-northeast tilting in southern hemisphere, is necessary for transporting westerly momentum from mid-to-high latitudes to the tropical region for maintaining the equatorial superrotation \citep{Vallis_2006}. This up-gradient transport of angular momentum by waves and eddies is necessary for maintaining the equatorial superrotation, which is called as the Hide's theorem \citep{Hide_1969}.

\begin{figure*}
  \centering
  \includegraphics[width=0.70\textwidth]{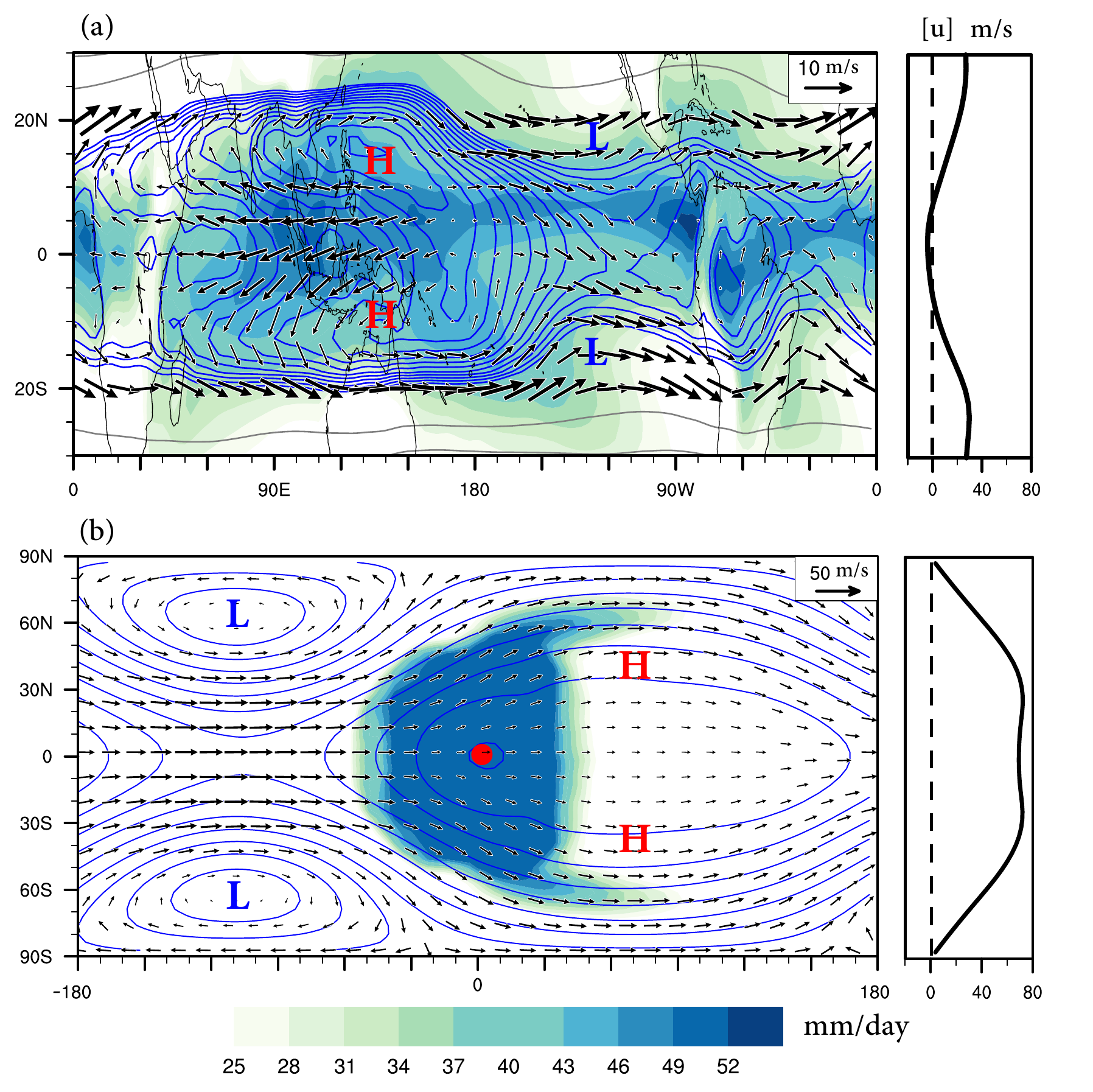}
  \caption{Comparisons between the tropical atmospheric circulation on Earth (a) and the global atmospheric circulation on a simulated tidally locked terrestrial planet. Black vectors are the winds at 150 hPa in (a) and 200 hPa in (b); contour lines are the corresponding geopotential height; and color shading is the annual-mean precipitation. On the right of the panels, the corresponding zonal-mean zonal winds are shown. H: high-pressure center, and L: low-pressure center. (a) is for the tropical region between 30$^\circ$S and 30$^\circ$N, and (b) is for the global due to the larger Rossby deformation radius for the tidally locked planet. For (b), the steller flux is 1200 W\,m$^{-2}$, the planetary rotation period (=\,orbital period) is 37 Earth days, the surface air pressure is 1.0 bar, and atmospheric CO$_2$ concentration is 300~ppmv in the simulation. Note the zonal shifts of the high- and low-pressure centers between (a) and (b).}
  \label{fig1_comparison}
\end{figure*}

Recent studies showed that the superrotation can also influence the phase and shape of the Rossby and Kelvin waves through horizontal advection (or called ``Doppler Shift''), forming a tropical wave--mean flow interaction problem between the atmospheric waves and the background flow \citep{Tsai_2014,Hammond_2018}. The Doppler shift can be viewed in the comparison between Earth and tidally locked planets, as shown in Fig.~\ref{fig1_comparison}. On Earth, latent heat release during deep convection over the warm pool of the west Pacific Ocean causes convergence in the low troposphere and divergence in the high troposphere and induces Rossby and Kelvin waves in the atmosphere \citep{Matsuno_1966,Gill_1980}. The Rossby waves propagate to the west side and the Kelvin waves propagate to the east side. In the upper troposphere, the spatial pattern is characterized by a pair of anticyclonic Rossby gyres (i.e., high-pressure centers) symmetrically located about the equator in the west of the maximum latent heating and by an equatorial Kelvin wave in the east (Fig.~\ref{fig1_comparison}(a), see also \cite{Gill_1980} and \cite{Dima_2005}). The spatial size of the gyres is mainly determined by the equatorial Rossby deformation radius \citep{Vallis_2006}. In the simulations of tidally locked terrestrial planet, similar phenomenon can be found but in a global scale due to the slow rotation of the planet and thereby a much larger equatorial Rossby deformation radius, as shown in Fig.~\ref{fig1_comparison}(b). Importantly, the phase of the waves is quite different. The anticyclonic Rossby gyres are on the east of the substellar point while the cyclonic Rossby gyres are on the west, meaning a large phase shift, compared to the observations on Earth and to the results of \cite{Gill_1980} with a zero mean flow. Previous studies on the phase shift have examined a limit range of parameters \citep{Tsai_2014,Hammond_2018}. In this work, we systemically examine the phase shift feature and find that the degree of the phase shift is not unlimited and one upper limit of $+\pi$ and one lower limit of $-\pi$ exist in the system.

Note that the phase shift of planetary waves emphasized in this study is different from the phase offset (or called phase shift or hotspot shift) of the phase curve in observations, although there are certain connections between them. The former is the zonal shift of the crests and troughs of Rossby and Kelvin waves, relative to their original locations under no mean flow. The latter is the zonal shift of the disk-integrated thermal
energy of the planets measured by a distant observer, relative to the distribution of the stellar radiation on the planet. Partially, the degree of the phase offset can be influenced by the waves. Other processes can also have important effects on the phase curve, such as cloud distribution, haze, water vapor concentration, and oceanic heat transport from the dayside to the nightside if the planets are habitable and have ocean(s).

Furthermore, \cite{Arnold_2012}, \cite{Tsai_2014} and \cite{Herbert_2020} shown that the amplitude of the waves is a nonlinear function of the speed of the mean flow and that the system exhibits resonant state, within which the amplitude of the waves (as well as their associated horizontal momentum transports) reaches a peak. The resonance occurs as the velocity of the mean flow approaches to the phase speed of free Rossby wave but in opposite sign or it approaches to the phase speed of free Kelvin wave but also in opposite sign \citep{Herbert_2020}. The resonance behaviour is similar to the singularity of Rossby waves forced by a westerly wind over topography on Earth as addressed in \cite{Held_1983} and in Chapter 5.7.2 of \cite{Holton_2013}. In this study, we re-investigate the resonance behaviour; based on energy budget of the system, we will more clearly show the underlying physical mechanism (see section {\ref{session_resonances}} below).

The goal of this study is improving our understanding of the atmospheric circulation on tidally locked planets mainly based on a 2D idealized shallow water model. The advantage of this work is that our analytic solution includes a zonal jet and meanwhile allows unequal radiative timescale and drag timescale, whereas previous studies did not. The structure of this paper is as follows. Section~\ref{session_methods} describes the shallow water equations and their analytic solutions. Section~\ref{session_results} shows the phase shift of the Rossby and Kelvin waves as a function of the strength of the mean flow (section \ref{session_phase_shift}), the resonance between the waves and the mean flow under which the amplitude of the waves reaches maximum (section \ref{session_resonances}), and the result of a 3D global circulation experiment (section \ref{session_GCM_results}). Section~\ref{session_summary} is the summary and discussions.


\section{Methods}\label{session_methods}

\subsection{The Linear Shallow Water Model}\label{session_equations}

A 2D linear shallow water model in equatorial $\beta$-plane with Matsuno-Gill-type forcing \citep{Matsuno_1966,Gill_1980,Showman_2011,Heng_Workman_2014,Penn_2017} is used in this study. The model is 1.5-layer shallow water system, including an active upper layer of constant density that represents the free troposphere and underlying a quiescent layer representing the lower troposphere. Atmospheric flows are forced by steady heating and cooling (corresponding to mass source and sink in the model) and damped by radiative relaxation and linear friction. The heating and cooling are assumed small enough for linear theory to apply. For easy to obtain analytic solutions, a uniform mean flow ($U$), representing an atmospheric superrotation, is specified in the model, same as that employed in \cite{Phlips_1987}, \cite{Arnold_2012} and \cite{Tsai_2014}. To enable analytic solutions, nonlinear advection terms are not considered, except the zonal advection by the specified mean flow. These lead to a linear system for the flow in the upper layer:

\begin{equation}
\frac{\partial u}{\partial t}+U\frac{\partial u}{\partial x}-\beta yv+g\frac{\partial\eta}{\partial x}=-\frac{u}{\tau_{drag}}, \label{eq1}
\end{equation}

\begin{equation}
\frac{\partial v}{\partial t}+U\frac{\partial v}{\partial x}+\beta yu+g\frac{\partial\eta}{\partial y}=-\frac{v}{\tau_{drag}}, \label{eq2}
\end{equation}

\begin{equation}
\frac{\partial\eta}{\partial t}+U\frac{\partial\eta}{\partial x}+H\left(\frac{\partial u}{\partial x}+\frac{\partial v}{\partial y}\right)=S(x,y)-\frac{\eta}{\tau_{rad}}, \label{eq3}
\end{equation}
where $t$ is the time, $x$ is the eastward distance, $y$ is the northward distance, $u$ is the zonal velocity, $v$ is the meridional velocity, $\beta=df/dy$ is the gradient of the Coriolis parameter with the northward distance, $\eta$ is the height anomaly from the reference fluid height in the absence of forcing ($H$), and $g$ is the reduced gravity (i.e., gravity times the fractional density difference between the layers $\Delta\rho/\rho$). Momentum dissipation is represented by  Rayleigh friction with a drag timescale of $\tau_{drag}$. Heating and cooling in the atmosphere are represented by mass source and sink, respectively, writing as $S\,=\,(h_{eq}\,-\,H)/\tau_{rad}$, where $\tau_{rad}$ is the radiative relaxation timescale (i.e., the timescale for the atmosphere to reach local radiative equilibrium) and $h_{eq}$ is the 2D radiative-equilibrium height. These linear equations imply that the height anomaly is much smaller than the average fluid thickness ($\eta\,\ll\,H$). Momentum exchange between the upper layer and the lower layer \citep{Showman_2010} is neglected in the model. The momentum exchange term is important in generating the superrotation in a nonlinear shallow water system \citep{Showman_2010}, but it is unimportant in a linear shallow water system when investigating the behavior of waves rather than the superrotation. In math, the momentum exchange term is quadratic and does not appear in the linear shallow water equations \citep{Showman_2011}. Similar to \cite{Showman_2011}, we have ignored the term here in order to obtain analytical solution, and moreover the superrotation wind is specified in our model.



\begin{figure*}
  \centering
  \includegraphics[width=1.0\textwidth]{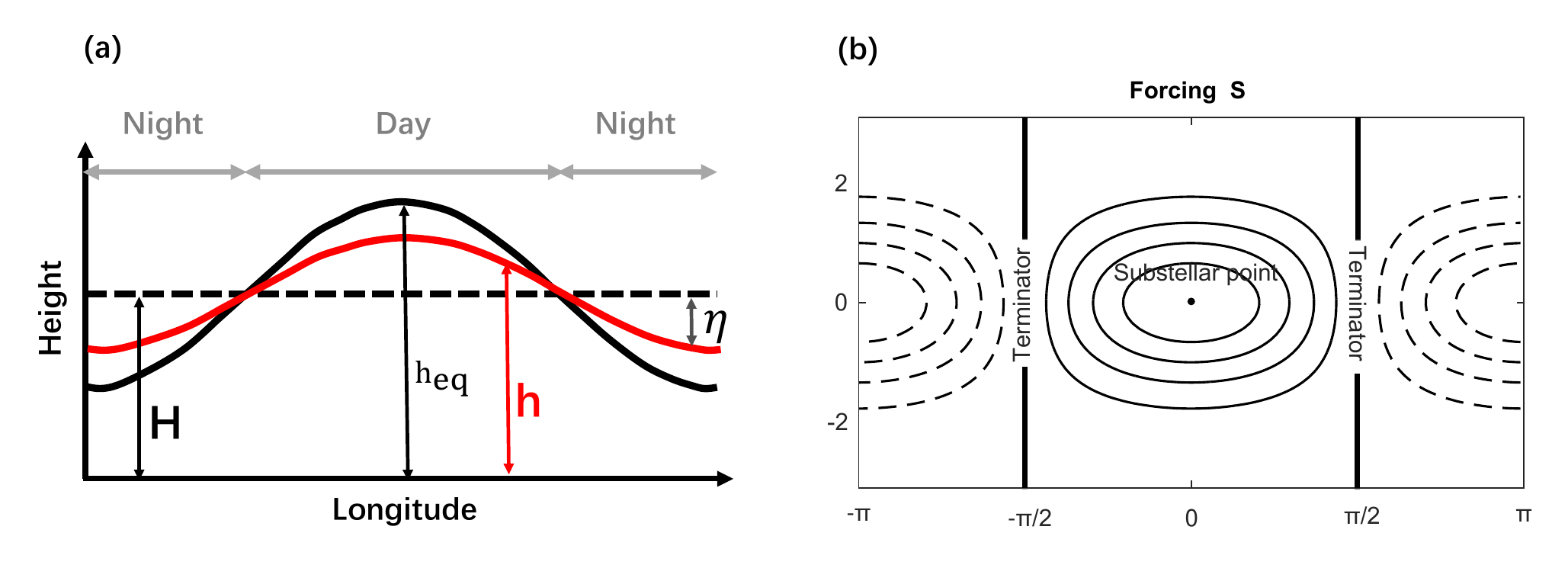}
  \caption{(a) Diagram of the shallow water model. The reference height ($H$), the radiative equilibrium height ($h_{eq}$), the actual height ($h$), and the height anomaly ($\eta$) are labelled. (b) Contours of the forcing field in the system ($S$ in Equation~(\ref{eq7})) with solid contours for mass source on the dayside and dashed contours for mass sink on the nightside. The substellar point is at 0 latitude and 0 longitude.}
  \label{fig2_SWM}
\end{figure*}

We nondimensionalize the equations with a velocity scale of  $c=\sqrt{gH}$ corresponding to the speed of gravity wave (such as the nondispersive Kelvin wave), a length scale of $L_R=(c/\beta)^{1/2}$ corresponding to the equatorial Rossby deformation radius, and a timescale of $T=(c\beta)^{-1/2}$ corresponding to the time for the gravity wave to propagate  throughout the distance of $L_R$. The height scale is equal to $H$, the dissipation and radiative timescales are nondimensionalized with $T$, and $S$ with $H/T$. These characteristic scales are the same as those used in  \cite{Matsuno_1966} and \cite{Showman_2011}. For steady solutions under forcing and damping, the equations become

\begin{equation}
U^*\frac{\partial u^*}{\partial x^*}+\frac{\partial\eta^*}{\partial x^*}-y^*v^*=-\frac{u^*}{\tau^{*}_{drag}}, \label{eq4}
\end{equation}

\begin{equation}
U^*\frac{\partial v^*}{\partial x^*}+\frac{\partial\eta^*}{\partial y^*}+y^*u^*=-\frac{v^*}{\tau^{*}_{drag}}, \label{eq5}
\end{equation}

\begin{equation}
U^*\frac{\partial\eta^*}{\partial x^*}+\left(\frac{\partial u^*}{\partial x^*}+\frac{\partial v^*}{\partial y^*}\right)=S^*(x^*,y^*)-\frac{\eta^*}{\tau^{*}_{rad}}, \label{eq6}
\end{equation}
within which all variables with stars are nondimensional. For brevity, we omit stars in the following text. The model now  contains three main important parameters: $U$, $\tau_{drag}$, and $\tau_{rad}$. Our goal in Section~\ref{session_results} below is to determine the dependence of the phase shift and the resonance on these three parameters.

For 1:1 tidally locked planets, the forcing is a sinusoidal function in longitude and a parabolic cylinder function in latitude, in order to approximately mimic the mass source on the dayside and the mass sink on the nightside, same as that employed in previous studies \cite[such as][]{Showman_2010,Showman_2011}, written as
\begin{equation}
S(x,y)=\frac{\Delta h_{eq}}{\tau_{rad}}\sum_{n}\psi_{n}(y)e^{ikx}, \label{eq7}
\end{equation}
where $\psi_{n}(y)$ are parabolic cylinder functions, subscript $n$ is the order of the functions \citep{Abramowitz_1965}, and $\Delta h_{eq}$ represents the dimensionless strength of the radiative-equilibrium force. The substellar point is at $kx=0$ and the antistellar point is at $kx=\pi$, thus positive $e^{ikx}$ corresponds to mass source on the dayside and negative corresponds to mass sink on the nightside, as shown in Fig.~\ref{fig2_SWM}(b). Due to the hemisphere-scale mass source/sink on tidally locked planets, we consider $\psi_{0}(y)$ only in the following solutions. Note that a flat nightside is more realistic according to the spatial pattern of the stellar radiation \citep{Perez-Becker_2013,Showman_2013,Komacek_2016,Zhang_2017}, but it is not used in this work; this is due to that obtained analytic solution is not easy under a flat nightside when the radiative timescale and the dissipation timescale are unequal. Furthermore, the waves do not exhibit essential differences when comparing a flat nightside to a sinusoidal function on the nightside, although the detailed spatial pattern and amplitude of the waves do vary \citep{Perez-Becker_2013}.


The model used here is similar to that in \cite{Showman_2011} but a mean flow is included, so that our model is able to consider the effect of the mean flow on the waves. \cite{Tsai_2014} also considered the effect of a uniform mean flow on the waves in both 2D and 3D linear models. But, the model used here is relatively easier to uncover the underlying mechanisms. Moreover, in the work of \cite{Tsai_2014}, the radiative timescale is assumed to be equal to the dissipation timescale (i.e., $\tau_{rad}=\tau_{drag}$) due to the constraint in their method of obtaining analytic solutions, while in our model this assumption is unnecessary, so that we can discuss the separate effects of varying $\tau_{rad}$ and varying $\tau_{drag}$. \cite{Hammond_2018} introduces a nonuniform jet $U(y)$ in their 2D shallow water model, which is more realistic and can better match their results up with 3D atmospheric circulation simulations in horizontal height field, but it was inconvenient for obtaining the results analytically; a pseudo-spectral method is required for a nonuniform jet \citep{Hammond_2018}.

\subsection{Analytic Solutions}\label{session_solution}

To solve Eqs.~(\ref{eq4})--(\ref{eq6}), we follow the method used in  \cite{Showman_2011} but including the mean jet of $U$. Due to the longitudinally sinusoidal form of the forcing (see Fig.~\ref{fig2_SWM}(b)), all the variables would  have the same form in longitude:
\begin{equation}
\left[u,v,\eta,S\right]=\left[u(y),v(y),\eta(y),S(y)\right]e^{ikx}. \label{eq8}
\end{equation}

The goal is to solve for the unknowns of $u$, $v$, and $\eta$, under given $S(y)$. In order to solve the equations, it is convenient to introduce several variables: $\alpha \equiv \tau_{drag}^{-1}$, $\gamma \equiv \tau_{rad}^{-1}$, $\tilde{\alpha}\equiv\alpha+ikU$, $\tilde{\gamma}\equiv\gamma+ikU$, $q(y)=\sqrt{\tilde{\gamma}}\eta(y)+\sqrt{\tilde{\alpha}}u(y)$, and $r(y)=\sqrt{\tilde{\gamma}}\eta(y)-\sqrt{\tilde{\alpha}}u(y)$, where $q(y)$ and $r(y)$ are complex variables due to the introduction of the zonal-mean jet $U$. So, we can convert Eqs.~(\ref{eq4})--(\ref{eq6}) to
three equivalent equations for $q(y)$, $r(y)$, and $v(y)$:

\begin{equation}
ikq(y)+\sqrt{\tilde{\alpha}}\frac{dv(y)}{dy}=\sqrt{\tilde{\gamma}}yv(y)+\sqrt{\tilde{\alpha}}S(y) -\sqrt{\tilde{\alpha}\tilde{\gamma}}q(y), \label{eq9}
\end{equation}

\begin{equation}
ikr(y)-\sqrt{\tilde{\alpha}}\frac{dv(y)}{dy}=\sqrt{\tilde{\gamma}}yv(y)-\sqrt{\tilde{\alpha}}S(y) +\sqrt{\tilde{\alpha}\tilde{\gamma}}r(y), \label{eq10}
\end{equation}

\begin{equation}
yq(y)-yr(y)+\sqrt{\frac{\tilde{\alpha}}{\tilde{\gamma}}}\frac{dq(y)}{dy}+ \sqrt{\frac{\tilde{\alpha}}{\tilde{\gamma}}}\frac{dr(y)}{dy}=-2\tilde{\alpha}^{3/2}v(y). \label{eq11}
\end{equation}
The eigenfunctions of $q(y)$, $r(y)$, $v(y)$, and $S(y)$ are parabolic cylinder functions:
\begin{equation}
\left[q(y),r(y),v(y),S(y)\right]=\sum_{n}\left[\hat{q}_n,\hat{r}_n,\hat{v}_n,\hat{S}_n\right]\psi_{n}(y), \label{eq12}
\end{equation}
where $\hat{S}_0=\gamma\Delta h_{eq}$. The parabolic cylinder functions ($\psi_{n}(y)$) and their recursion relations are

\begin{equation}
\psi_{n}(y)=\mathrm{exp}\left(-\frac{y^2}{2\mathcal{P}^2}\right)H_n\left(\frac{y}{\mathcal{P}}\right), \label{eq13}
\end{equation}

\begin{equation}
\frac{d\psi_n}{dy}=\frac{2n\psi_{n-1}}{\mathcal{P}}-\frac{y\psi_n}{\mathcal{P}^2}, \label{eq14}
\end{equation}

\begin{equation}
\frac{d\psi_n}{dy}=-\frac{\psi_{n+1}}{\mathcal{P}}+\frac{y\psi_n}{\mathcal{P}^2}, \label{eq15}
\end{equation}
where $H_n(y)$ is the Hermitian polynomial and $\mathcal{P}\equiv(\tilde{\alpha}/\tilde{\gamma})^{1/4}$ \citep{Abramowitz_1965}. Substituting  Eq.~(\ref{eq12}) into Eqs.~(\ref{eq9})--(\ref{eq11}) and using the recursion relations, we obtain three series equations:

\begin{equation}
ik\hat{q}_n+\sqrt{\tilde{\alpha}\tilde{\gamma}}\hat{q}_n-\left(\tilde{\alpha}\tilde{\gamma}\right)^{1/4} \hat{v}_{n-1}=\sqrt{\tilde{\alpha}}\hat{S}_n, \label{eq16}
\end{equation}

\begin{equation}
ik\hat{r}_n-\sqrt{\tilde{\alpha}\tilde{\gamma}}\hat{r}_n-2(n+1)\left(\tilde{\alpha}\tilde{\gamma}\right)^{1/4} \hat{v}_{n+1}=-\sqrt{\tilde{\alpha}}\hat{S}_n, \label{eq17}
\end{equation}

\begin{equation}
2(n+1)\hat{q}_{n+1}-\hat{r}_{n-1}=-2\tilde{\alpha}^{3/2}\mathcal{P}^{-1}\hat{v}_n, \label{eq18}
\end{equation}
where the subscript $n\geq0$, and the terms will be zero if the subscript $n<0$. The given forcing is $\hat{S}_0$, and the order $n$ is no more than 2, as mentioned in \cite{Gill_1980} and \cite{Showman_2011}. For $n=0$, Eq.~(\ref{eq16}) implies

\begin{equation}
\hat{q}_0=\frac{\sqrt{\tilde{\alpha}}\left(\sqrt{\tilde{\alpha}\tilde{\gamma}}-ik\right)}{ \tilde{\alpha}\tilde{\gamma}+k^2}\hat{S}_0. \label{eq19}
\end{equation}

\begin{figure*}
  \centering
  \includegraphics[width=0.9\textwidth]{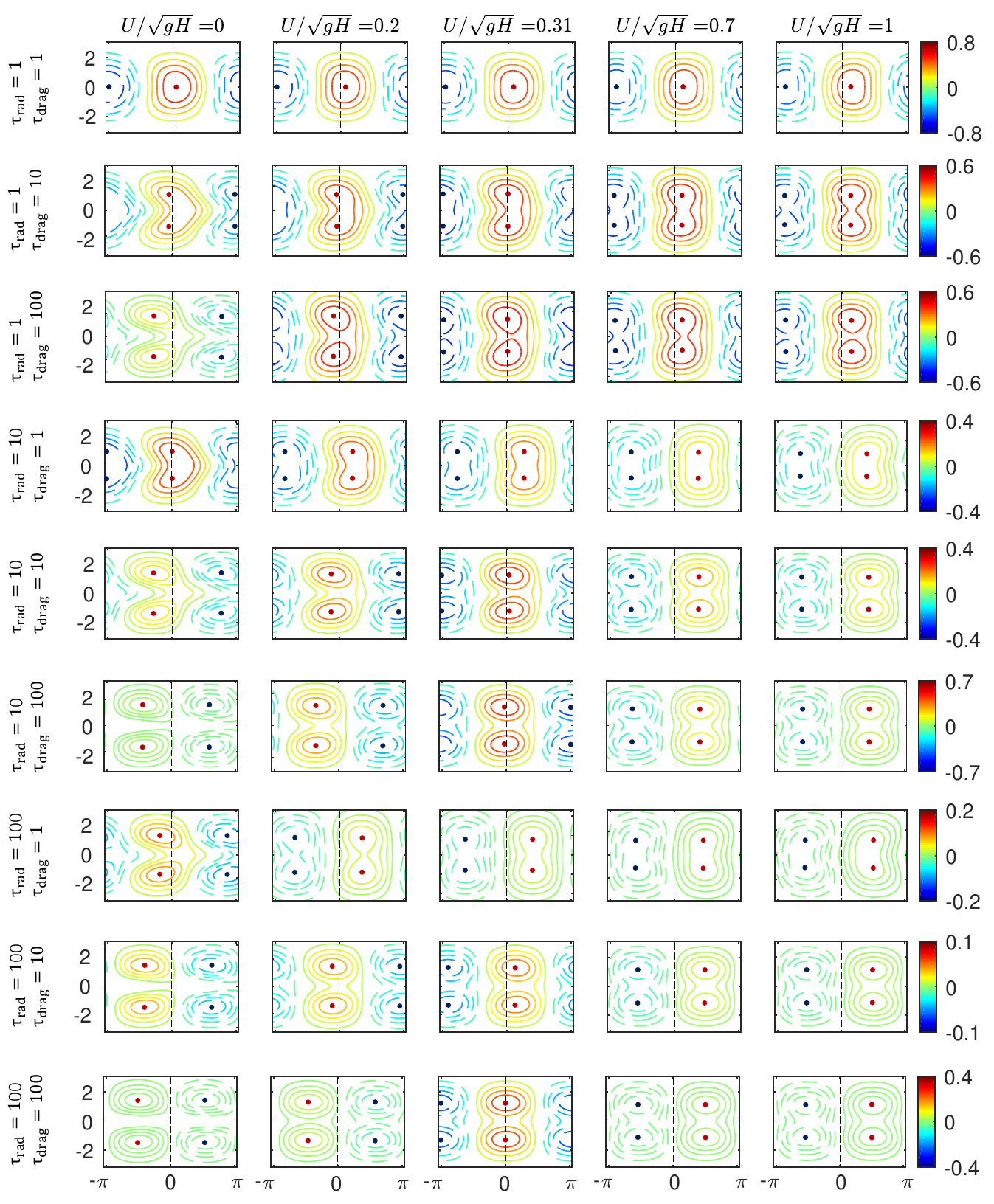}
  \caption{Examples of the height anomaly field ($\eta$) in the linear shallow water equations (Eqs.~(\ref{eq4})--(\ref{eq6})) under different timescales and different strengths of the westerly mean flow ($U\geq0$). From left to right, the magnitude of the mean flow is 0, 0.2, 0.31, 0.7, and 1.0 of the gravity wave speed ($\sqrt{gH}$), respectively. Nine combinations for three radiation timescales ($\tau_{rad}$\,=\,1, 10, and 100) and three drag timescales ($\tau_{drag}$\,=\,1, 10, and 100) are shown. In order to more clear show the phase of the waves, zonal-mean values have been subtracted. The substellar point lies at the center of each panel. Note that the colorbar for each row is different, due to the fact that the radiative and drag timescales significantly influence the magnitudes of the Rossby and Kelvin waves. For each row, the contour interval is 0.2 of the maximum value of the corresponding colorbar. The $y$-axis of each panel is $\mathrm{Real}\left(\tilde{\gamma}/\tilde{\alpha}\right)^{1/4}y$.}
  \label{figure_sum_pattern1}
\end{figure*}

Substituting $n=2$, $n=0$, and $n=1$ into Eqs.~(\ref{eq16})-(\ref{eq18}), we obtain:

\begin{equation}
ik\hat{q}_2+\sqrt{\tilde{\alpha}\tilde{\gamma}}\hat{q}_2-\left(\tilde{\alpha}\tilde{\gamma}\right)^{1/4} \hat{v}_1=0, \label{eq20}
\end{equation}

\begin{equation}
ik\hat{r}_0-\sqrt{\tilde{\alpha}\tilde{\gamma}}\hat{r}_0-2\left(\tilde{\alpha}\tilde{\gamma}\right)^{1/4} \hat{v}_1=-\sqrt{\tilde{\alpha}}\hat{S}_0, \label{eq21}
\end{equation}

\begin{equation}
4\hat{q}_2-\hat{r}_0=-2\tilde{\alpha}^{3/2}\mathcal{P}^{-1}\hat{v}_1. \label{eq22}
\end{equation}
Eliminating $\hat{r}_0$ and $\hat{v}_1$, we obtain

\begin{equation}
\hat{q}_2=\frac{\tilde{\alpha}^{3/2}k^2+3\tilde{\alpha}\tilde{\gamma}^{1/2}+\tilde{\alpha}^{5/2}\tilde{\gamma}+ik\tilde{\alpha}^{1/2}} {2\left[\left(\tilde{\alpha}k^2+3\tilde{\alpha}^{1/2}\tilde{\gamma}^{1/2}+\tilde{\alpha}^2\tilde{\gamma}\right)^2+k^2\right]} \hat{S}_0. \label{eq23}
\end{equation}
From Eq.~(\ref{eq20}) and the analytic solution of $\hat{q}_2$, we get an analytic solution for $\hat{v}_1$:

\begin{equation}
\hat{v}_1=\frac{\sqrt{\tilde{\alpha}\tilde{\gamma}}+ik}{\left(\tilde{\alpha}\tilde{\gamma}\right)^{1/4}} \hat{q}_2. \label{eq24}
\end{equation}
Combining Eqs.~(\ref{eq22}) and~(\ref{eq24}), we obtain
$\hat{r}_0=\left(4+2\tilde{\alpha}^{3/2}\sqrt{\tilde{\gamma}}+2ik\tilde{\alpha}\right)\hat{q}_2.$ Note that $\hat{r}_2$ is equal to zero. The final solution is:
\begin{eqnarray}
    \hat{\eta}&=&\frac{\hat{q}_0\psi_0+\hat{q}_2\psi_2+ \hat{r}_0\psi_0}{2\sqrt{\tilde{\gamma}}}e^{ikx}
    =\hat{S}_0\left[\sqrt{\frac{\tilde{\alpha}}{\tilde{\gamma}}}\frac{\sqrt{\tilde{\alpha}\tilde{\gamma}}-ik}{2(\tilde{\alpha}\tilde{\gamma}+k^2)}e^{ikx}\right]\mathrm{exp}\left(-\sqrt{\frac{\tilde{\gamma}}{\tilde{\alpha}}}\frac{y^2}{2}\right) \nonumber \\
    &+&\hat{S}_0\left[\frac{\tilde{\alpha}^{3/2}k^2+3\tilde{\alpha}\tilde{\gamma}^{1/2}+\tilde{\alpha}^{5/2}\tilde{\gamma}+ik\tilde{\alpha}^{1/2}} {4\sqrt{\tilde{\gamma}}\left[\left(\tilde{\alpha}k^2+3\tilde{\alpha}^{1/2}\tilde{\gamma}^{1/2}+\tilde{\alpha}^2\tilde{\gamma}\right)^2+k^2\right]}e^{ikx}\right]\left(4\sqrt{\frac{\tilde{\gamma}}{\tilde{\alpha}}}y^2-2\right)\mathrm{exp}\left(-\sqrt{\frac{\tilde{\gamma}}{\tilde{\alpha}}}\frac{y^2}{2}\right) \nonumber \\
    &+&\hat{S}_0\left[\left(2\tilde{\gamma}^{-1/2}+\tilde{\alpha}^{3/2}+ik\tilde{\alpha}\tilde{\gamma}^{-1/2}\right)\frac{\tilde{\alpha}^{3/2}k^2+3\tilde{\alpha}\tilde{\gamma}^{1/2}+\tilde{\alpha}^{5/2}\tilde{\gamma}+ik\tilde{\alpha}^{1/2}}{2\left[\left(\tilde{\alpha}k^2+3\tilde{\alpha}^{1/2}\tilde{\gamma}^{1/2}+\tilde{\alpha}^2\tilde{\gamma}\right)^2+k^2\right]}e^{ikx}\right]\mathrm{exp}\left(-\sqrt{\frac{\tilde{\gamma}}{\tilde{\alpha}}}\frac{y^2}{2}\right), \label{eq25}
\end{eqnarray}

\begin{eqnarray}
    \hat{u}&=&\frac{\hat{q}_0\psi_0+\hat{q}_2\psi_2- \hat{r}_0\psi_0}{2\sqrt{\tilde{\alpha}}}e^{ikx}
    =\hat{S}_0\left[\frac{\sqrt{\tilde{\alpha}\tilde{\gamma}}-ik}{2(\tilde{\alpha}\tilde{\gamma}+k^2)}e^{ikx}\right]\mathrm{exp}\left(-\sqrt{\frac{\tilde{\gamma}}{\tilde{\alpha}}}\frac{y^2}{2}\right) \nonumber \\
    &+&\hat{S}_0\left[\frac{\tilde{\alpha}^{3/2}k^2+3\tilde{\alpha}\tilde{\gamma}^{1/2}+\tilde{\alpha}^{5/2}\tilde{\gamma}+ik\tilde{\alpha}^{1/2}} {4\sqrt{\tilde{\alpha}}\left[\left(\tilde{\alpha}k^2+3\tilde{\alpha}^{1/2}\tilde{\gamma}^{1/2}+\tilde{\alpha}^2\tilde{\gamma}\right)^2+k^2\right]}e^{ikx}\right]\left(4\sqrt{\frac{\tilde{\gamma}}{\tilde{\alpha}}}y^2-2\right)\mathrm{exp}\left(-\sqrt{\frac{\tilde{\gamma}}{\tilde{\alpha}}}\frac{y^2}{2}\right) \nonumber \\
    &-&\hat{S}_0\left[\left(2\tilde{\alpha}^{-1/2}+\tilde{\alpha}\tilde{\gamma}^{1/2}+ik\tilde{\alpha}^{1/2}\right)\frac{\tilde{\alpha}^{3/2}k^2+3\tilde{\alpha}\tilde{\gamma}^{1/2}+\tilde{\alpha}^{5/2}\tilde{\gamma}+ik\tilde{\alpha}^{1/2}}{2\left[\left(\tilde{\alpha}k^2+3\tilde{\alpha}^{1/2}\tilde{\gamma}^{1/2}+\tilde{\alpha}^2\tilde{\gamma}\right)^2+k^2\right]}e^{ikx}\right]\mathrm{exp}\left(-\sqrt{\frac{\tilde{\gamma}}{\tilde{\alpha}}}\frac{y^2}{2}\right),\label{eq26}
\end{eqnarray}
and
\begin{eqnarray}
    \hat{v}&=&\hat{v}_1\psi_1e^{ikx} \nonumber \\
    &=&\hat{S}_0\left[\frac{(\sqrt{\tilde{\alpha}\tilde{\gamma}}+ik)(\tilde{\alpha}^{3/2}k^2+3\tilde{\alpha}\tilde{\gamma}^{1/2}+\tilde{\alpha}^{5/2}\tilde{\gamma}+ik\tilde{\alpha}^{1/2})}{(\tilde{\alpha}\tilde{\gamma})^{1/4}\left[\left(\tilde{\alpha}k^2+3\tilde{\alpha}^{1/2}\tilde{\gamma}^{1/2}+\tilde{\alpha}^2\tilde{\gamma}\right)^2+k^2\right]}e^{ikx}\right](\tilde{\gamma}/\tilde{\alpha})^{1/4}y\mathrm{exp}\left(-\sqrt{\frac{\tilde{\gamma}}{\tilde{\alpha}}}\frac{y^2}{2}\right),  \label{eq27}
\end{eqnarray}
where $\hat{\eta}$, $\hat{u}$, and $\hat{v}$ are complex quantities and $\eta$, $u$, and $v$ are the real parts of them.

\begin{figure*}
  \centering
  \includegraphics[width=1.0\textwidth]{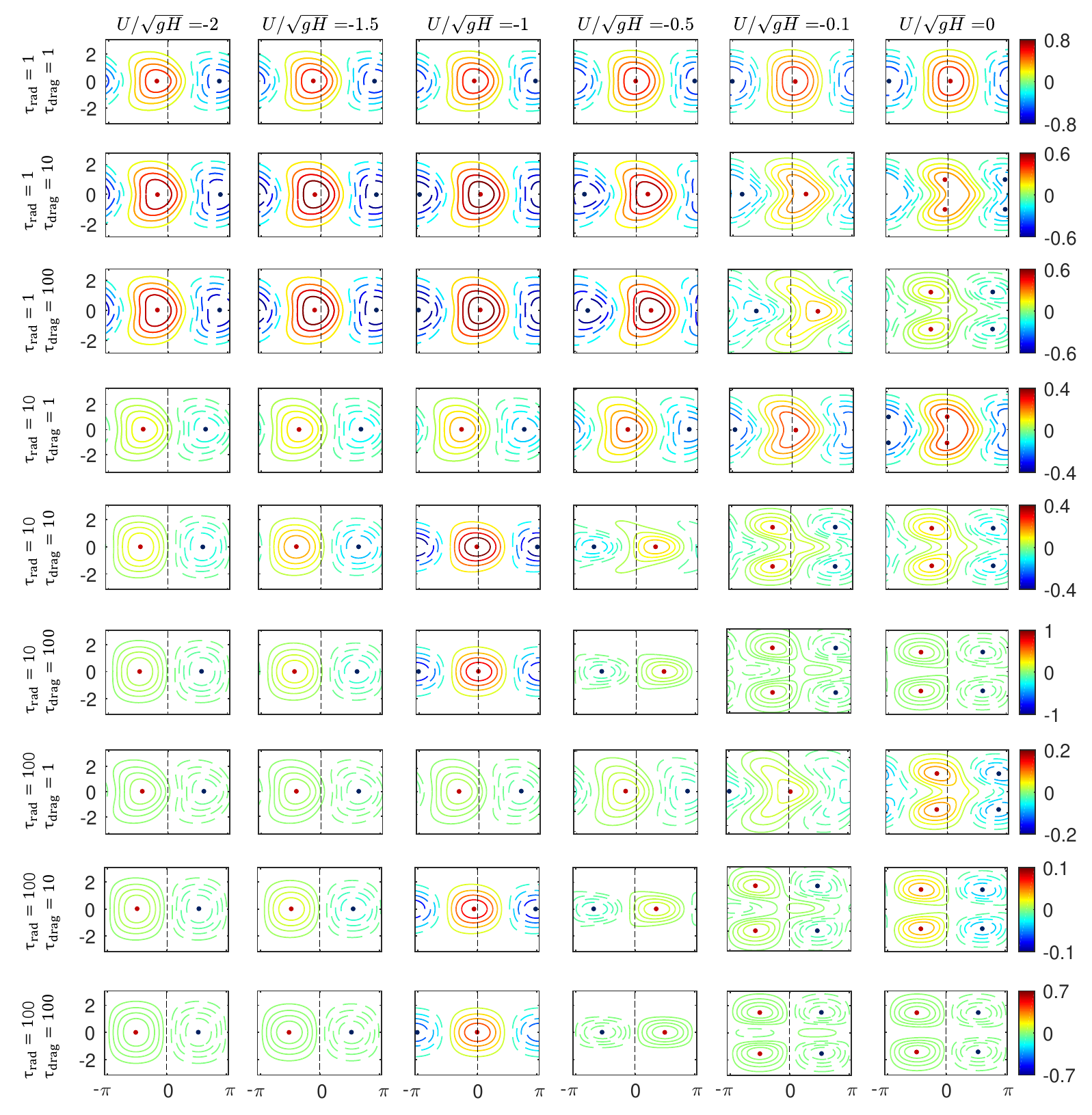}
  \caption{Same as Figure~\ref{figure_sum_pattern1} but under an easterly mean flow ($U\leq0$). From left to right, the magnitude of the mean flow is $-$2.0, $-$1.5, $-$1.0, $-$0.5, $-$0.1 and 0 of the gravity wave speed ($\sqrt{gH}$), respectively. The rightest column is the same as the leftest column in Fig.~\ref{figure_sum_pattern1} for comparison.}
  \label{figure_sum_pattern2}
\end{figure*}

The solution can be decomposed to Kelvin and Rossby components  \citep{Gill_1980, Vallis_2006}. For the Kelvin wave, its meridional velocity ($v$) is equal to zero, and its $u$ and $\eta$ satisfy
$\eta_{K}=\frac{\hat{q}_0\psi_0}{2\sqrt{\tilde{\gamma}}}e^{ikx}$
and $u_{K}=\frac{\hat{q}_0\psi_0}{2\sqrt{\tilde{\alpha}}}e^{ikx}$, respectively. The Rossby waves are $\eta_{R}=\frac{\hat{q}_2\psi_2+\hat{r}_0\psi_0}{2\sqrt{\tilde{\gamma}}}e^{ikx}$, $u_{R}=\frac{\hat{q}_2\psi_2-\hat{r}_0\psi_0}{2\sqrt{\tilde{\alpha}}}e^{ikx}$, and $v_{R}=\hat{v}_1\psi_1e^{ikx}$. Below, Figs.~\ref{figure_sum_pattern1} and \ref{figure_sum_pattern2} show the typical solutions under different values of $\tau_{rad}$, $\tau_{drag}$, and $U$.

\section{Results}\label{session_results}

\subsection{Phase Shifts of Rossby and Kelvin Waves Under Mean Flow}\label{session_phase_shift}

The atmospheric circulations under zero and non-zero mean flows are shown in Fig.~\ref{figure_sum_pattern1}. When $U$\,=\,$0$, the spatial pattern of the circulation results from the superposition of equatorial Kelvin wave that propagates to the east and off-equatorial Rossby waves that propagate to the west (see the leftmost panels). When the drag or cooling effect is small, an outstanding feature of the circulation pattern is the high pressure center (i.e., the local maximum of the height anomaly ($\eta$); zonal-mean values have been subtracted) and low pressure center (i.e., the local minimum of $\eta$) located off the equator. In the vicinity of the equator, the flow is directing from the dayside to the nightside, caused by the surface inclination associated with the mass sources and sinks. In higher latitudes, due to the effect of the Coriolis force, the flow field is approximately in geostrophical balance with the pressure (surface elevation) field, within which anticyclonic or cyclonic flow fields are established where high- or low-pressure cells are located \citep{Matsuno_1966}. The flow off the equator is connected to the flow on the equator through converging or diverging motions towards or from the equator at the end or beginning of each cell. When the drag and cooling effect is strong (top left corner of Fig.~\ref{figure_sum_pattern1}), the spatial pattern is analogous to the source/sink. However, wave pattern still exhibits slight northwest-southeast tilting in the northern hemisphere and southwest-northeast tilting in the southern hemisphere although the zonal propagation of Rossby and Kelvin waves is nearly inhibited. This is due to that the drag and Coriolis forces are comparable, and a three-way horizontal force balance between those two forces and pressure-gradient force leads to the tilting \citep{Showman_2013}. Note that inertia gravity waves are not included in the solutions. This is due to the facts that we are looking for a steady solution under stationary forcing and that the frequencies of the Rossby and Kelvin waves are much smaller than those of inertia gravity waves \citep{Matsuno_1966}.

When $U$\,=\,$0$, the eddy high pressure center (marked with red dot in each panel) is around the substellar point when the radiation timescale ($\tau_{rad}$) or the drag timescale ($\tau_{drag}$) is short, or on the west of the substellar point when $\tau_{rad}$ and $\tau_{drag}$ are intermediate or long. When $\tau_{rad}$ (or $\tau_{drag}$) is short, the effect of radiation relaxation (or friction damping) is strong, so that the waves are unable to effectively propagate zonally, the height anomaly field is more analogous to the radiative equilibrium height, and therefore the high pressure center is close to the substellar point and the low pressure is close to the antistellar point. When $\tau_{rad}$ and $\tau_{drag}$ are long, the eddy height anomaly field is dominated by off-equatorial Rossby waves, which effectively propagates towards west, so that the high pressure centers (corresponding to wave crest) are on the west of the substellar point and the low pressure centers (corresponding to wave trough) are on the east of the substellar point (the leftmost panels in Fig.~\ref{figure_sum_pattern1}, see also Fig. 3 in \cite{Showman_2011}).

\begin{figure*}
  \centering
  \includegraphics[width=0.9\textwidth]{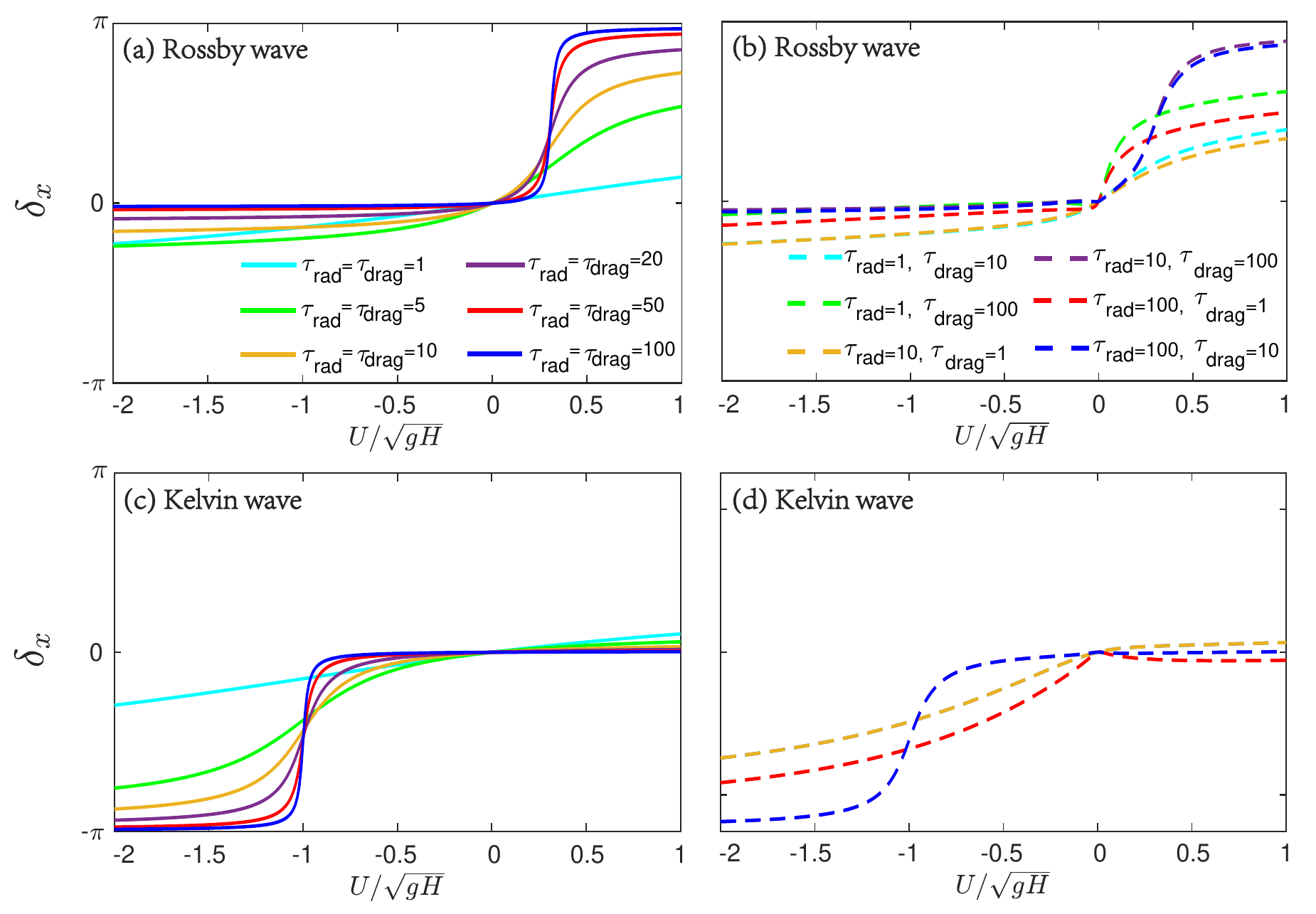}
  \caption{The degree of the phase shift of Rossby waves (a--b) and Kelvin waves (c--d) as a function of the strength of the mean flow. Left panels show the cases within which the radiation timescale being equal to the drag timescale ($\tau_{rad}=\tau_{drag}=$1, 5, 10, 20, 50, and 100), and right panels show the cases of unequal $\tau_{rad}$ and $\tau_{drag}$ of 1, 10, and 100. In panel (d), the line of  $\tau_{rad}=1$ \& $\tau_{drag}=10$ is overlapped by the line of  $\tau_{rad}=10$ \& $\tau_{drag}=1$, and the same for the $\tau_{rad}=1$ \& $\tau_{drag}=100$ and $\tau_{rad}=100$ \& $\tau_{drag}=1$ lines and for the $\tau_{rad}=10$ \& $\tau_{drag}=100$ and $\tau_{rad}=100$ \& $\tau_{drag}=10$ lines.}
  \label{figure_phase_shift}
\end{figure*}

As $U$\,$>$\,$0$, a clear response is that both the high and low pressure centers shift towards the east, meaning an eastward shift in the phase of the waves (Fig.~\ref{figure_sum_pattern1}). The eastward phase shift is mainly from the Rossby waves while the Kelvin wave component remains almost stationary (figure not shown), similar to that found in Fig.~7 of \cite{Tsai_2014}. This is due to that the phase speed of the Kelvin wave ($\sqrt{gH}$) is always faster than the mean flow, so it is less affected. The magnitude of the phase shift depends on the magnitude of $U$ and on the values of $\tau_{rad}$ and $\tau_{drag}$ (Fig. \ref{figure_phase_shift}(a \& b)). For a given $U$, the magnitude of the phase shift is smaller when $\tau_{rad}$ or $\tau_{drag}$ is shorter (i.e., the radiative relaxation or the drag friction is stronger), and vice versa. Mathematically speaking, Eqs.~(\ref{eq4}-\ref{eq6}) shows that when the radiation relaxation term and the friction term are strong, the effect of the advection terms would be relatively smaller.

For $U$\,$<$\,$0$, both the high and low pressure centers shift towards the west (Fig.~\ref{figure_sum_pattern2}), meaning a westward phase shift of the waves. The westward phase shift is mainly from the Kelvin wave while the Rossby wave component remains almost stationary (figure not shown). The magnitude of the westward phase shift also depends on the magnitude of $U$ and on the values of $\tau_{rad}$ and $\tau_{drag}$, and it is relatively small (large) when $\tau_{rad}$ or $\tau_{drag}$ is shorter (longer) (Fig. \ref{figure_phase_shift}(c \& d)).

Figure~\ref{figure_phase_shift} shows that under given $\tau_{rad}$ and $\tau_{drag}$, the degree of the phase shift is a monotonic, nonlinear increasing function of $U$ and more importantly there are limits for the phase shift, $+\pi$ for $U>0$ and $-\pi$ for $U<0$. When $|U|$ is increased, the transition from a small phase shift to the upper limits of $\pm\pi$ under large $\tau_{rad}$ and $\tau_{drag}$ is sharper than that under small $\tau_{rad}$ and $\tau_{drag}$. The limits of $\pm\pi$ in the phase shift can be explained based on the energy balance of the whole system. For Eqs.~(\ref{eq1})--(\ref{eq3}), the corresponding global energy equation is given by
\begin{eqnarray}
\frac{\partial}{\partial t}\int\int(K_E+P_E)dxdy
=\int\int(\eta S)dxdy-\int\int\left(\frac{2K_E}{\tau_{drag}}+\frac{2P_E}{\tau_{rad}}\right)dxdy, \label{eq28}
\end{eqnarray}
where $K_E\equiv(u^2+v^2)/2$ and $P_E\equiv\eta^2/2$ represent nondimensional kinetic energy and potential energy, respectively. This equation is obtained through multiplying Eq.~(\ref{eq1}) by $u$, multiplying Eq.~(\ref{eq2}) by $v$, multiplying Eq.~(\ref{eq3}) by $\eta$, and then spatially integrating the equations under periodic boundary condition in the $x$ direction and zero normal velocity condition at the north and south side boundaries.

\begin{figure*}
  \centering
  \includegraphics[width=0.80\textwidth]{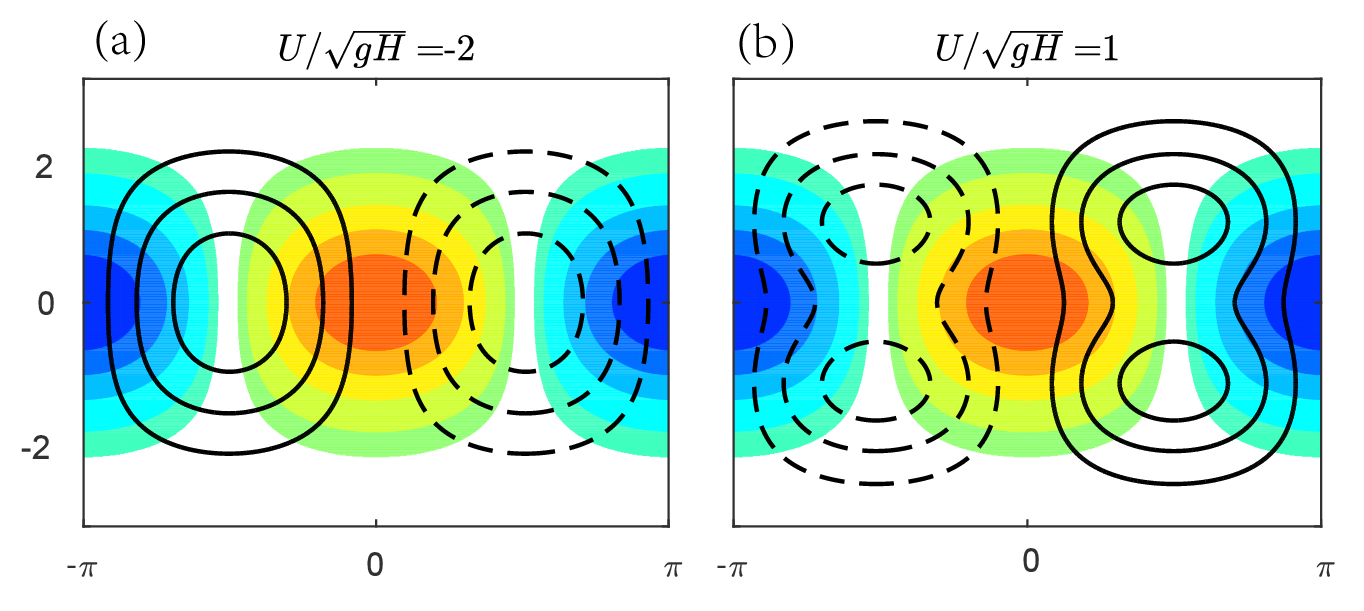}
  \caption{Schematic diagram for the limits of the phase shift: $-\pi$ under an easterly mean flow (a) and $+\pi$ under a westerly mean flow (b). Red color and blue color show the mass source and sink, respectively; solid lines show high pressure cells, and dashed lines show low pressure cells. In both panels, the correlation between the height anomaly ($\eta$) and the forcing ($S$) in global integration is close to zero.}
  \label{figure_upper_limit}
\end{figure*}

The first term on the right side of Eq.~(\ref{eq28}) is the correlation between the height anomaly ($\eta$) and the forcing ($S$), which represents the exchange between the source energy and the kinetic and potential energy of the waves. The second term is the damping and friction effects in the system. Because the second term is negative everywhere, the first term must be positive at least in global mean and thereby a positive correlation between $\eta$ and $S$ is required. From Fig.~\ref{figure_upper_limit}(a), it is clear to see that under a westward mean flow of such as $U=-2\sqrt{gH}$, the phase shift reaches the lower limit of $-\pi$; this is because a further westward shift of $\eta$ will cause a zero or negative integration of $\eta S$ and a steady state cannot reach. Similarly, under an eastward mean flow of such as $U=+\sqrt{gH}$, the phase shift reaches the upper limit of $+\pi$, because a further eastward shift of $\eta$ will cause a zero or negative integration of $\eta S$ (Fig.~\ref{figure_upper_limit}(b)). These results mean that the center of the Kelvin wave cannot shift beyond the west terminator and the center of the Rossby waves cannot shift beyond the east terminator.

The two limits in the phase shift can also be understood based on a solution at the equator. At $y=0$, the shallow water equations (Eqs.~(\ref{eq4}-\ref{eq6}) and Eq.~(\ref{eq8})) reduce to
\begin{equation}
\tilde{\alpha}u^0+ik\eta^0=0, \label{eq29}
\end{equation}
\begin{equation}
\tilde{\gamma}\eta^0+iku^0=\gamma S_0-v^0_y, \label{eq30}
\end{equation}
where the superscript 0 means quantities at $y=0$, and the subscript $y$ represents the partial derivative with respect to the latitude. The solution of the height field is
\begin{equation}
\eta^0=\frac{\alpha^2\gamma+\alpha k^2+\gamma k^2U^2+ikU\left(k^2-\alpha^2-k^2U^2\right)}{\left(\alpha\gamma-k^2U^2+k^2\right)^2+k^2U^2\left(\alpha+\gamma\right)^2}\left[\gamma S_0-2\left(\frac{\gamma+ikU}{\alpha+ikU}\right)^{1/4}\hat{v}_1\right], \label{eq31}
\end{equation}
where $\hat{v}_1$ is the same as that in Eq.~(\ref{eq24}). When $\alpha$ and $\gamma$ are small (i.e., $\tau_{drag}$ and $\tau_{rad}$ are large) and $U$ is large enough ($kU\gg\alpha,\gamma$), $\hat{v}_1$ is proportion to $\tilde{\alpha}^{-2}$ and hence $U^{-2}$ (ref. Eq.~(\ref{eq27})), so it tends to be close to $0$. Thereby, the value of $\eta^0$ in Eq.~(\ref{eq31}) is close to $-i\gamma S_0/kU$. This means that the
absolute value of the maximum phase deviation from $S_0$ is $\pi/2$ and the upper limit of the phase shift is $+\pi$. The same principle applies to the lower limit.


\subsection{Resonance between Planetary Waves and Mean Flow}\label{session_resonances}

\begin{figure*}
  \centering
  \includegraphics[width=0.90\textwidth]{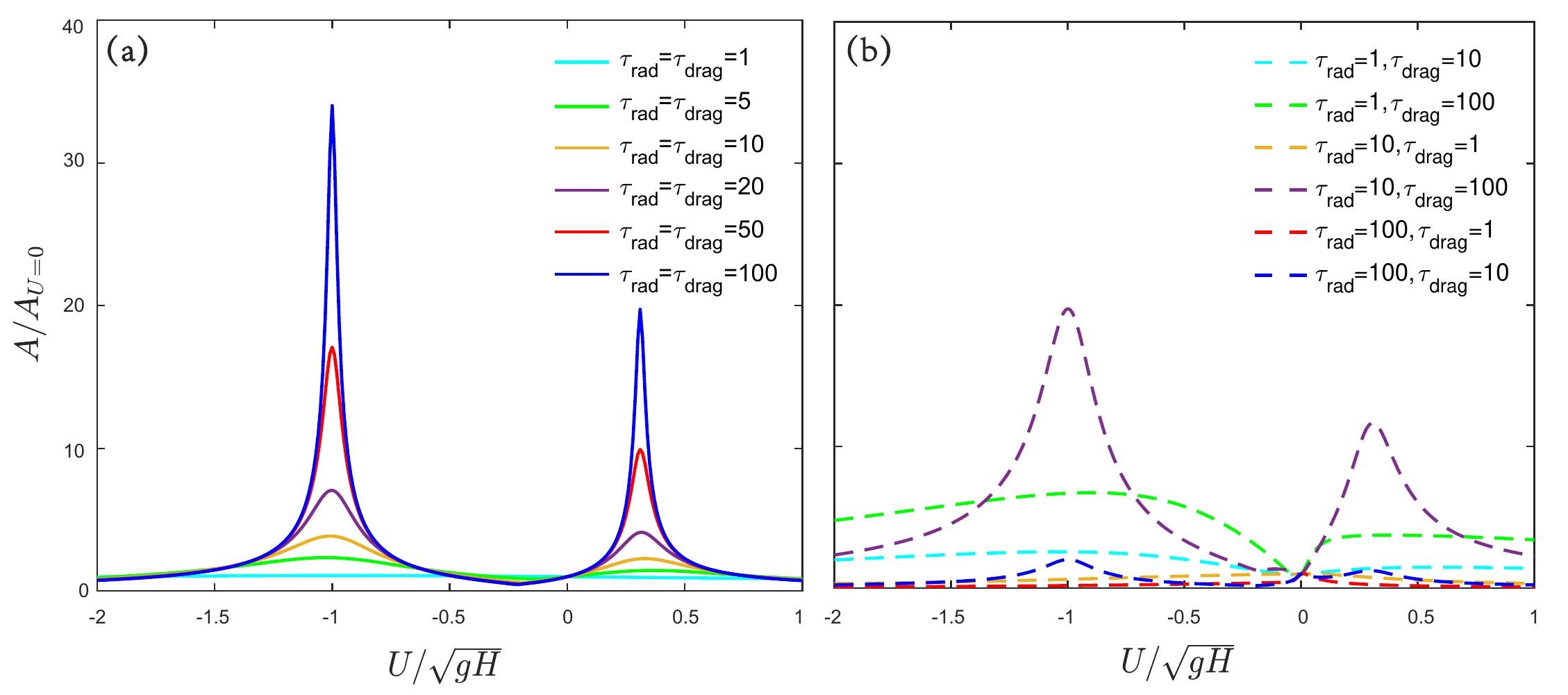}
  \caption{Response of the wave amplitude to the mean flow, defined as the ratio of the height amplitude under a non-zero mean flow to that under a zero mean flow. Different lines represent the combinations of different radiative and drag timescales, same as that in Fig.~\ref{figure_phase_shift}. Left panel: $\tau_{rad}=\tau_{drag}$, and right panel: $\tau_{rad}\neq\tau_{drag}$.}
  \label{figure_amplitude}
\end{figure*}

Besides of the phase shift of planetary waves, the mean flow can also influence the amplitude of the waves. As shown in Fig.~\ref{figure_amplitude}, the amplitude of the waves (compared to the condition under a zero mean flow) exhibits a resonance behaviour. For $U>0$, the wave amplitude reaches a peak when the value of $U$ approaches to the westward phase speed of the Rossby wave, which is $\sqrt{gH}/3$. For $U<0$, the wave amplitude also reaches a peak when the absolute value of $U$ approaches to the eastward phase speed of the  Kelvin wave, which is $\sqrt{gH}$. When $\tau_{rad}$ or $\tau_{drag}$ is small, the resonance behaviour is weak or absence due to the strong relaxing or damping effect, while when $\tau_{rad}$ and $\tau_{drag}$ are large, the resonance is more significant. This resonance phenomena has also
been found in the previous studies of \cite{Arnold_2012}, \cite{Tsai_2014}, and \cite{Herbert_2020} but only for identical $\tau_{rad}$ and $\tau_{drag}$.

\begin{figure*}
  \centering
  \includegraphics[width=0.80\textwidth]{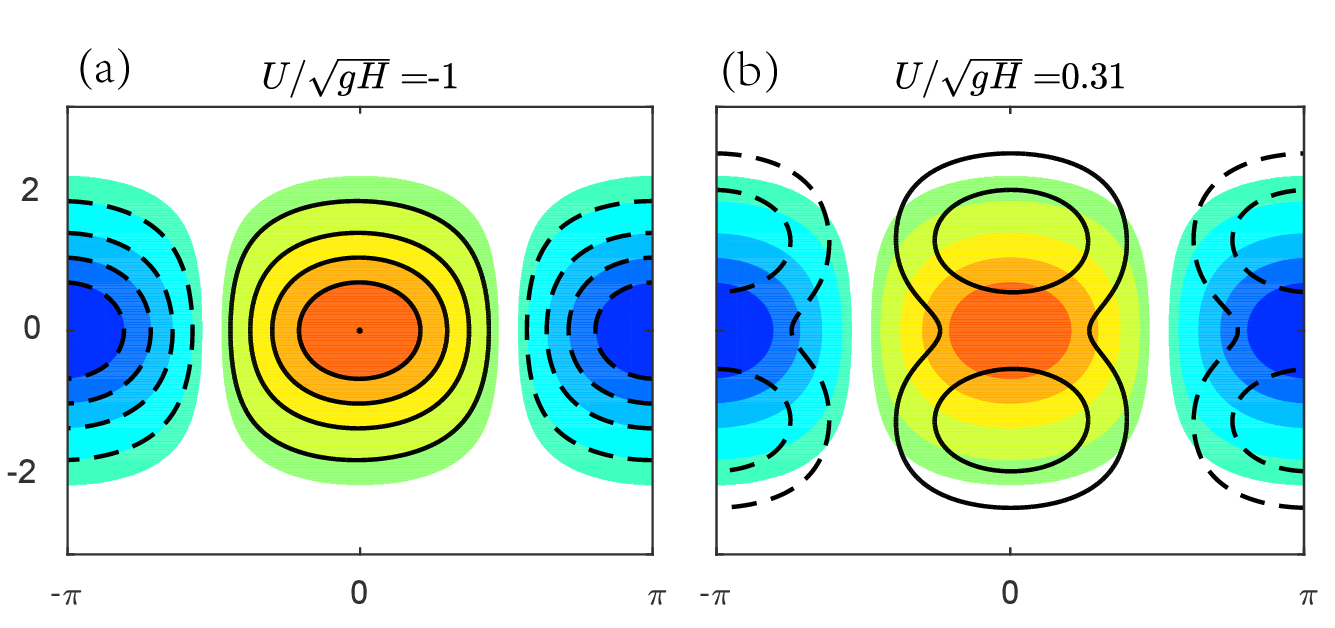}
  \caption{Schematic diagram for the wave--jet resonance. Color shading shows the mass source (red) and the mass sink (blue); black lines show high pressure cell (solid) and low pressure cell (dashed). In both panels, the correlation between the height anomaly ($\eta$) and the forcing ($S$) in global integration reaches a maximum value. The resonance occurs when the speed of an easterly mean flow is approximately equal to the eastward phase speed of Kelvin wave (a), or when the speed of an westerly mean flow is approximately equal to the westward phase speed of the Rossby wave (b).}
  \label{figure_resonance}
\end{figure*}

\begin{figure*}
  \centering
  \includegraphics[width=0.90\textwidth]{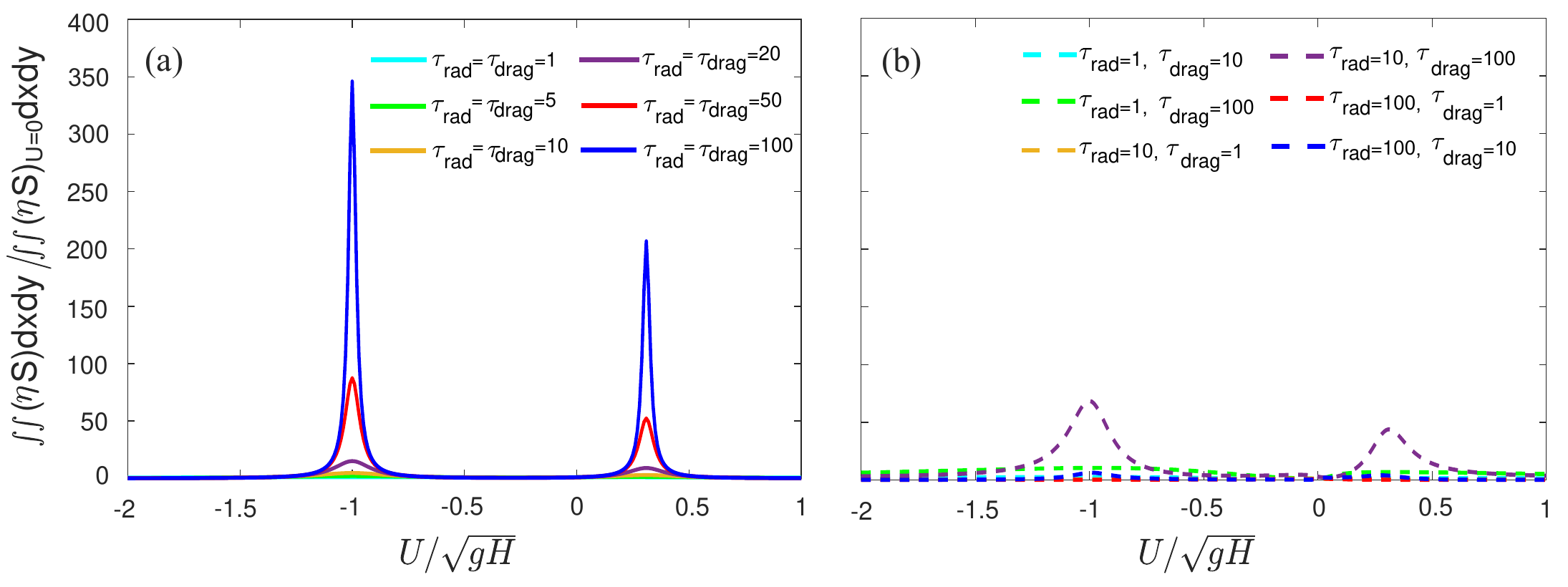}
  \caption{The energy exchange from the source to the kinetic and potential energy of the waves in the linear shallow water system. The lines show the ratio of  integrated energy exchange under $U\neq0$ to that under $U=0$. Line labels are the same as those in Fig.~\ref{figure_amplitude}.}
  \label{figure_energy_exchange}
\end{figure*}

The physical mechanism for the resonance can be interpreted based on the energy balance equation of Eq.~(\ref{eq28}). When the mean flow speed is equal to the phase speed of the Kelvin wave (left panel in Fig.~\ref{figure_resonance}) or that of the Rossby wave but in opposite sign (right panel in Fig.~\ref{figure_resonance}), the waves are right trapped in the mass source and sink regions, so that the correlation between the forcing ($S$) and the wave response ($\eta$) reaches a peak. This also means that the energy conversation from the source to the kinetic and potential energy of the waves reaches a maximum, as shown in Fig.~\ref{figure_energy_exchange}.

\begin{figure*}
  \centering
  \includegraphics[width=0.9\textwidth]{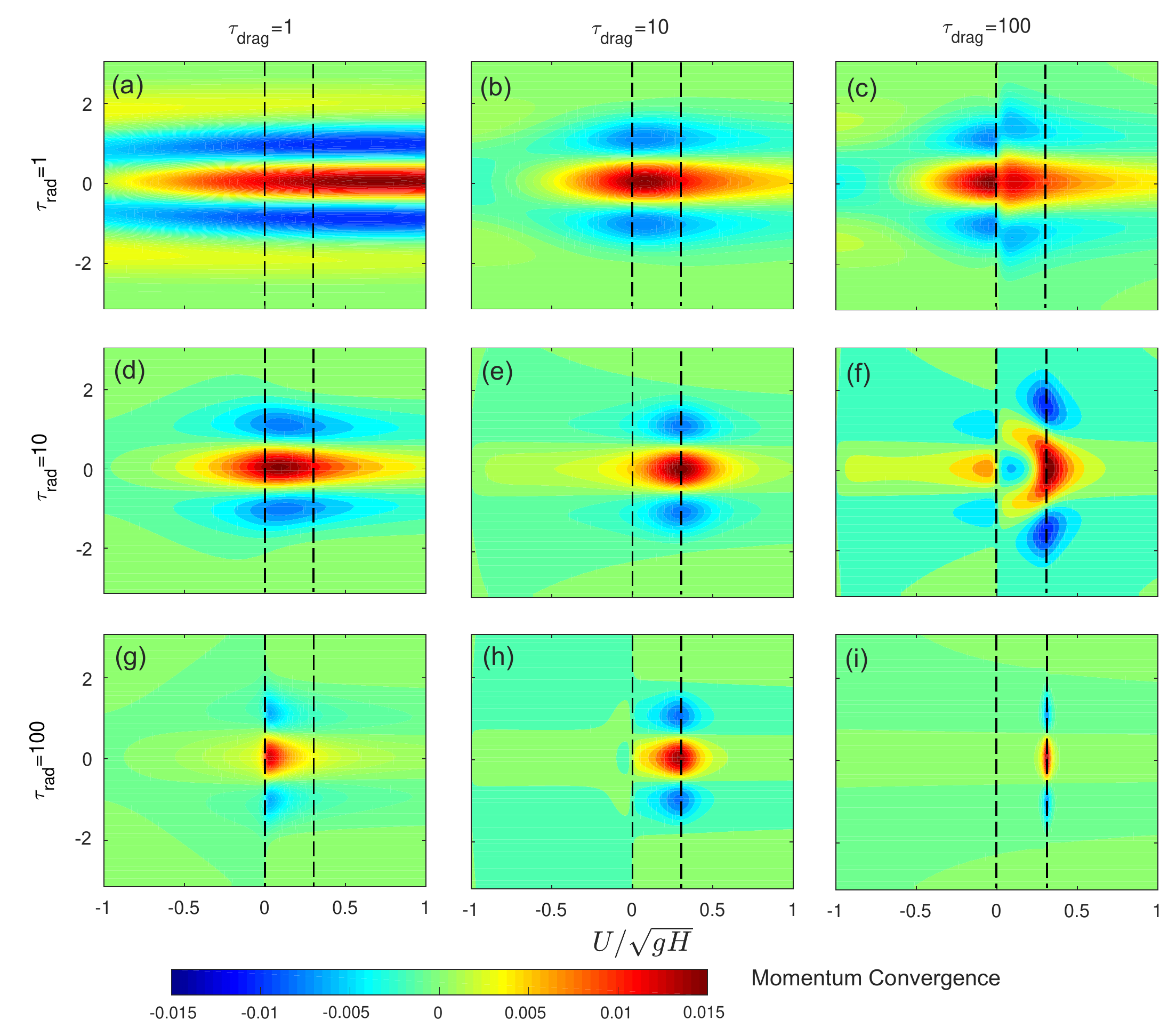}
  \caption{Horizontal eddy momentum flux convergence as functions of the strength of the mean flow ($x$ axis) and the latitude ($y$ axis). Positive and negative values represent accelerating and decelerating the westerly mean flow, respectively. Nine combinations for three radiation timescales ($\tau_{rad}$\,=\,1, 10, and 100) and three drag timescales ($\tau_{drag}$\,=\,1, 10, and 100) are shown. The two dashed lines represent $U/\sqrt{gH}$ being equal to 0 and 1/3, respectively. The momentum transport divergence is calculated using $-\frac{1}{\overline{h}}\frac{\partial{\overline{(hv)'u'}}}{\partial{y}}$, where the over-bar means the zonal average, while the prime means the anomaly after the zonal average \citep[see Eq.~(22) in][]{Showman_2011}. The $y$-axis of each panel is $\mathrm{Real}\left(\tilde{\gamma}/\tilde{\alpha}\right)^{1/4}y$.}
  \label{app}
\end{figure*}

The resonance behaviour can also be explained based on a simplified analytic solution. When $\tau_{rad}$ is equal to $\tau_{drag}$, the solution of $\eta$ reduces to:
\begin{equation}
\eta=\frac{S_0}{2\left[\alpha^2+k^2(U+1)^2\right]}(\alpha\cos(kx)+k(U+1)\sin (kx))\mathrm{exp}(-y^2/2), \label{eq32}
\end{equation}
for the Kelvin component, and
\begin{equation}
\eta=\frac{S_0}{6[\alpha^2+k^2(U-1/3)^2]}\left(\alpha\cos(kx)+k(U-1/3)\sin(kx)\right)\left(2y^2+1\right)\mathrm{exp}(-y^2/2), \label{eq33}
\end{equation}
for the Rossby component, similar to that found in \cite{Arnold_2012}. From these two equations, it is clear to see that the amplitude of the Kelvin wave reaches a maximum when the speed of the mean flow is approximately equal to $-\sqrt{gH}$, and the amplitude of the Rossby wave reaches a maximum when the speed of the mean flow is approximately equal to $\sqrt{gH}/3$, if all else being equal. The same explanation can be found in Equation (7) of \cite{Tsai_2014}: the wave responses are maximum when the Doppler-shifted frequency is equal to the frequency for free modes.

Note that because of the interaction between the Rossby and Kelvin waves, the mean flow speed for the occurring of the resonance is not exactly equal to the phase speed ($U/\sqrt{gH}\,=\,1/3$) but having a small deviation ($U/\sqrt{gH}\approx\,0.31$). Equation~(\ref{eq33}) shows that the amplitude of the Rossby wave is maximum at $U/\sqrt{gH}\,=\,1/3$, while Eq.~(\ref{eq32}) shows that the amplitude of the Kelvin wave decreases as the westerly wind speed  increases. Therefore, the coupled Rossby-Kelvin wave pattern resonates when $U/\sqrt{gH}$ is somewhat less than $1/3$.

The resonance behaviour as well as the phase shift of the waves can influence horizontal eddy momentum flux convergence, as shown in Fig.~\ref{app}. In this figure, a positive value means that the waves accelerate the mean flow and a negative value decelerate the mean flow. On the equator, the momentum flux convergence is positive as long as the mean flow is not too small and the corresponding value is negative in higher latitudes, indicating that zonal momentum is transported from the higher latitudes to the equatorial region. This acts to maintain the equatorial superrotation against friction and  dissipation \citep{Herbert_2020}. Markedly, there is a peak in the equatorial acceleration when the resonance between the Rossby waves and the mean flow occurs (panels (e), (f), (h), \& (i)). However, not all the acceleration peaks are occurring synchronously
 with the resonance. For example, when $\tau_{rad}$\,=\,1 and $\tau_{drag}$\,=\,1, the acceleration peak happens when the strength of the mean flow is equal to about 0.8 of the gravity wave speed (panel (a)), and when $\tau_{rad}$\,=\,1 and $\tau_{drag}$\,=\,100 (panel (c)) and $\tau_{rad}$\,=\,100 and $\tau_{drag}$\,=\,1 (panel (g)), the maximum acceleration happens when the strength of the mean flow is close to zero. This is due to the fact that the horizontal eddy momentum flux convergence is determined by the combined condition of the amplitude and the phase of the coupled Rossby-Kelvin waves; when the wave amplitude reaches a maximum, the tilt of the coupled waves may not be in an optimal condition for the  equator-ward momentum transport. Moreover, the resonance between the Kelvin wave and the mean flow does not imply a momentum convergence peak, as shown in Fig.~\ref{app}; this is due to that the tilt of the waves is very small or close to zero under this resonance (see the 3rd column of Fig.~\ref{figure_sum_pattern2}). These results suggest that a coupled system with both active waves and active mean flow is required in future work.


\subsection{Phase Shift and Resonance in a 3D AGCM Simulation}\label{session_GCM_results}

\begin{figure*}
  \centering
  \includegraphics[width=1.0\textwidth]{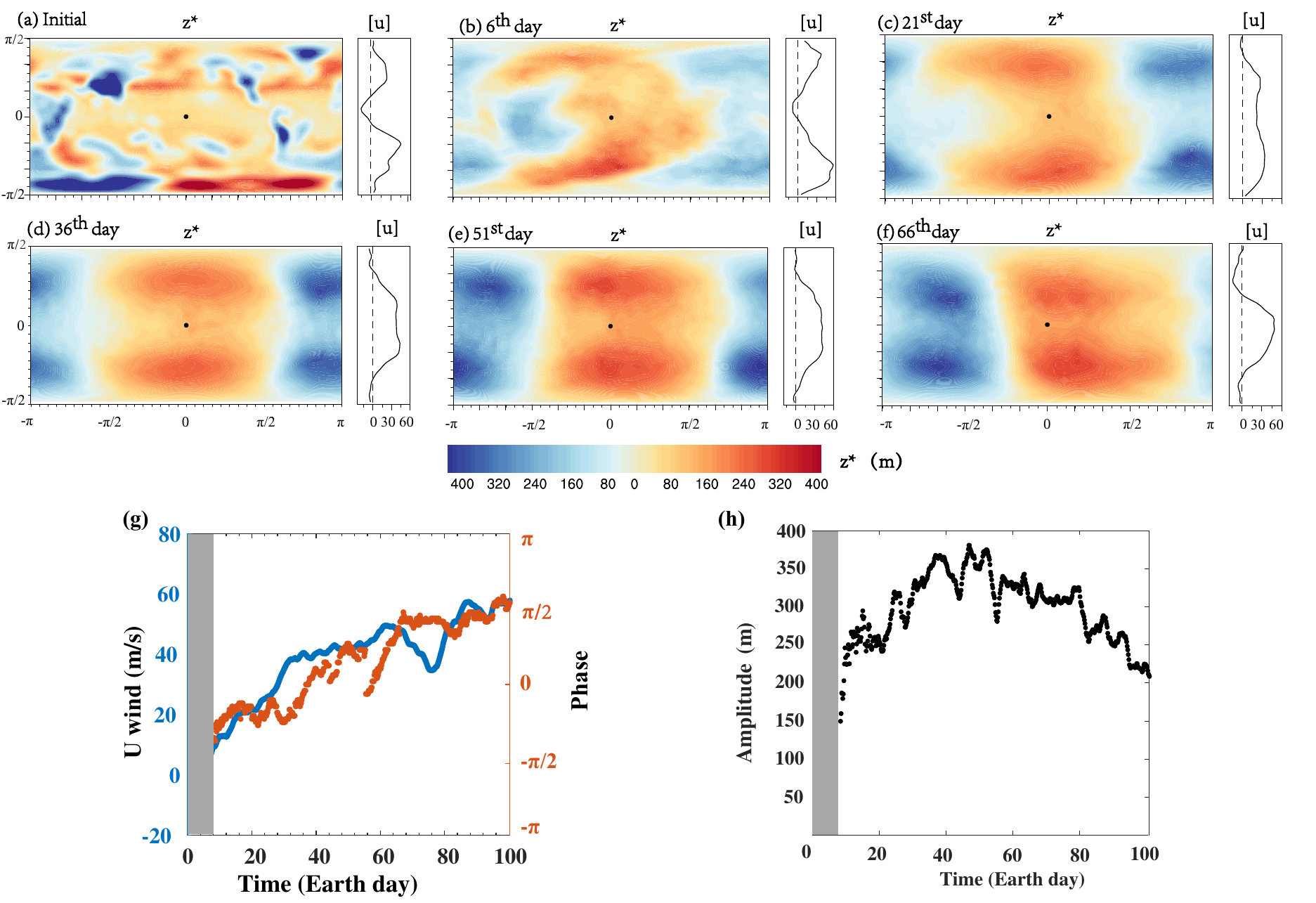}
  \caption{Phase shift of planetary waves and wave--jet resonance on a tidally locked terrestrial aqua-planet, simulated in the spin-up period of a 3D AGCM experiment. (a-f): eddy geopotential height at 200 hPa (color shading) and zonal-mean zonal winds (black line on the right of each panel, in units of m\,s$^{-1}$). (g): the speed of the equatorial mean flow (blue line) and the phase of the planetary wave (red line) as a function of time. (h): Amplitude of the planetary wave as a function of time. In (g--h), the first 8 Earth days are not shown because they are disorganized and should depend on the initial condition. Experiments with other rotation periods showed similar evolution processes (figures not shown).}
  \label{figure_GCM_spinup}
\end{figure*}

The phase shift of the planetary waves and the wave--jet resonance are also found in the spin-up period of a 3D atmospheric general circulation model (AGCM) experiment, as shown in Fig.~\ref{figure_GCM_spinup}. The experiment was performed using the global climate model CAM3, same as that used in \cite{Yang_2013}. The stellar flux was set to 1200~W\,m$^{-2}$, the star temperature is 3400 K, the rotation period (= orbital period) is 37 Earth days, the surface air pressure is 1.0 bar N$_2$, and atmospheric CO$_2$ concentration is 300~ppmv. The surface is covered by a 50-m slab ocean with no any continent. This experiment was initialized from a climate state similar to the present-day Earth.

As shown in Fig.~\ref{figure_GCM_spinup}, the wave pattern and atmospheric superrotation are established within about 100 Earth days. Initially, the geopotential height field at 200 hPa is roughly related to the surface land-sea distributions with high values over the oceans and low values over the continents (Fig.~\ref{figure_GCM_spinup}(a)). This is because the experiment was started from the summer atmospheric state of modern Earth, although the surface is set to be an aqua-planet in the simulation. In the 6th Earth day, a zonal-number one wave pattern is formed with high pressure centers in the west of the substellar point and with low pressure centers in the east of the substellar point. This wave pattern moves eastward gradually following the zonal flow, which exhibits an equatorial superrotation behaviour after about 10 Earth days. The amplitude of the waves also evolves with time and reaches a peak in about 50 Earth days (Fig.~\ref{figure_GCM_spinup}(h)) when the wave crest is right on the longitude line of the substellar point (Fig.~\ref{figure_GCM_spinup}(e)). After that, the waves move further east and the wave amplitude decreases. The wave crest never goes across the east terminator, likely implying the existence of an upper limit ($\pi$) for the phase shift. These results confirm that the phase shift of planetary waves and the wave--jet resonance on tidally locked planets do occur in a more realistic 3D AGCM simulation. In the 3D AGCM simulations of \cite{Arnold_2012}, \cite{Tsai_2014}, and \cite{Hammond_2018}, they showed similar phase shifts, but the resonance behaviour in 3D tidally locked AGCM simulations is the first time to be uncovered here.

\section{Summary and Discussions}\label{session_summary}
Equatorial superrotation is an important feature in the atmospheres of tidally locked hot jupiters and terrestrial planets. The maintenance of the superrotation is associated with equator-ward momentum transports by coupled Rossby-Kelvin waves, which are excited from the uneven distribution of stellar radiation between the permanent dayside and the permanent nightside. In this study, we investigate how the superrotation and in general a mean flow (either eastward or westward) influence the phase and the amplitude of the Rossby and Kelvin waves. We employ a 2D linear shallow water model because it is easier to uncover the underlying mechanisms, and we compare the results with a 3D AGCM simulation. Our method is similar to those used in the previous studies of \cite{Phlips_1987}, \cite{Arnold_2012}, \cite{Tsai_2014}, \cite{Hammond_2018}, and \cite{Herbert_2020}, but has some differences such as the unequal feature between the radiative timescale and the drag timescale. The main conclusions are:

   \begin{enumerate}
    \renewcommand{\labelenumi}{(\theenumi)}
      \item The mean flow is able to shift the phases of the Rossby and Kelvin waves. The degree of the phase shift is a nonlinear, monotonic function of the speed of the mean flow and can be influenced by the radiative and drag timescales. The phase shift has a upper limit of $+\pi$ and a lower limit of $-\pi$, which is constrained by the energy balance of the system.

      \item Resonance behaviour is robust in the tidally locked atmospheric system. The resonance occurs when the speed of an eastward mean flow is approximately equal to the westward phase speed of the Rossby wave or the speed of a westward mean flow is approximately equal to the eastward phase speed of the Kelvin wave. Under the resonance, the wave amplitude reaches a peak and the wave crest is trapped in the substellar region.

      \item Both the phase and the amplitude of the coupled Rossby-Kelvin  waves influence the equator-ward momentum transport and thereby affect the magnitude of the equatorial superrotation. When the resonance between the waves and the mean flow occurs, the momentum transport reaches a peak in some cases but not all.
   \end{enumerate}

 In this work, the mean flow is fixed, so the effect of the waves on the mean flow is not included in the shallow water model. Further work is required to develop a more realistic model within which both the two components are active. In the shallow water system, only the `tropical' Rossby and Kelvin waves on the Beta plane are included and the effects of `mid-latitude' phenomena, such as ageostrophic flow, baroclinic instability and eddy-driven jet \citep{Carone_2015,Kaspi_2015,Noda_2017,Haqq_Misra_2018,Lutsko_2018}, are not included in the model, so that the conclusions shown here are more suitable for planets having slow rotation rates, but the results are still applicable to the tropics of rapidly rotating planets. The effect of moisture convection and the upwelling and downwelling associated with the global-scale Walker circulation on tidally locked terrestrial planets are considered as simple mass source and sink in the model, whereas latent heat transport from the day side to the night side is also effective in reducing the day-to-night contrast, which can decrease the wave amplitudes \citep{Merlis_2010,Haqq_Misra_2018,Pierrehumbert_2018,Labont_2020}; as found in the moisture AGCM experiment shown in Fig.~\ref{figure_GCM_spinup}, the amplitude of the resonance is weaker than that predicted in the shallow water model (Fig.~\ref{figure_amplitude}). In this study, the jet is simply assumed as an uniform flow from the south pole to the north pole; in more complex models, the jet is trapped in the low latitudes and its width is mostly constrained by the scale of the Rossby deformation radius or the Rhines scale \citep{Showman_2013,Haqq_Misra_2018}. This can reduce the degrees of the phase shift and the amplitude resonance especially in the relatively high latitudes, as implied in the study of \cite{Hammond_2018}.

 Moreover, realistic radiative transfer, 3D structure, non-equilibrium chemistry, and detailed damping processes \citep[such as][]{Tuyl_1986,Wu_2001,Lin_2008,Showman_2013,Fujii_2017,Parmentier_Crossfield_2018,Shields_2019,Hammond_2020} can also influence the location and the amplitude of the waves, which  requires further studies. Finally, both of this study and previous studies focus on the waves and the equatorial superrotation in the atmosphere; whether these phenomena can also exist or not in the other fluid--the ocean on tidally locked planets \citep[such as][]{Hu_2014,Yang_2019,Del_2019} is only beginning to be understood.

\acknowledgments

We are grateful to the helpful discussions with Prof. Adam P. Showman (1968--2020), Daniel D.B. Koll, Yang Zhang, Zhaohua Wu, Yonggang Liu, Xianyu Tan, and Mark Hammond. J.Y. acknowledges support from the National Natural Science Foundation of China (NSFC) under grant 41675071.


\bibliography{sample63}{}

\begin{thebibliography}{}
\expandafter\ifx\csname natexlab\endcsname\relax\def\natexlab#1{#1}\fi
\providecommand{\url}[1]{\href{#1}{#1}}
\providecommand{\dodoi}[1]{doi:~\href{http://doi.org/#1}{\nolinkurl{#1}}}
\providecommand{\doeprint}[1]{\href{http://ascl.net/#1}{\nolinkurl{http://ascl.net/#1}}}
\providecommand{\doarXiv}[1]{\href{https://arxiv.org/abs/#1}{\nolinkurl{https://arxiv.org/abs/#1}}}

\bibitem[{Abramowitz \& Stegun(1965)}]{Abramowitz_1965}
Abramowitz, M., \& Stegun, I.~A. 1965, Handbook of Mathematical Functions with
  Formulas, Graphs, and Mathematical Tables (New York: Dover Publications Inc)

\bibitem[{Arnold {et~al.}(2012)Arnold, Tziperman, \& Farrell}]{Arnold_2012}
Arnold, N.~P., Tziperman, E., \& Farrell, B. 2012, Journal of the Atmospheric
  Sciences, 69, 626, \dodoi{10.1175/JAS-D-11-0136.1}

\bibitem[{Carone {et~al.}(2015)Carone, Keppens, \& Decin}]{Carone_2015}
Carone, L., Keppens, R., \& Decin, L. 2015, Monthly Notices of the Royal
  Astronomical Society, 453, 2412, \dodoi{10.1093/mnras/stv1752}

\bibitem[{Del~Genio {et~al.}(2019)Del~Genio, Way, Amundsen, Aleinov, Kelley,
  Kiang, \& Clune}]{Del_2019}
Del~Genio, A.~D., Way, M.~J., Amundsen, D.~S., {et~al.} 2019, Astrobiology, 19,
  99, \dodoi{10.1089/ast.2017.1760}

\bibitem[{Dima {et~al.}(2005)Dima, Wallace, \& Kraucunas}]{Dima_2005}
Dima, I.~M., Wallace, J.~M., \& Kraucunas, I. 2005, J. Atmos, 62, 2499,
  \dodoi{10.1175/JAS3486.1}

\bibitem[{Fujii {et~al.}(2017)Fujii, Del~Genio, \& Amundsen}]{Fujii_2017}
Fujii, Y., Del~Genio, A.~D., \& Amundsen, D.~S. 2017, The Astrophysical
  Journal, 848, 100, \dodoi{10.3847/1538-4357/aa8955}

\bibitem[{Gill(1980)}]{Gill_1980}
Gill, A.~E. 1980, Q. J. R. Meteorol. Soc., 106, 447,
  \dodoi{10.1002/qj.49710644905}

\bibitem[{Hammond \& Pierrehumbert(2018)}]{Hammond_2018}
Hammond, M., \& Pierrehumbert, R.~T. 2018, \apj, 869, 65,
  \dodoi{10.3847/1538-4357/aaec03}

\bibitem[{Hammond {et~al.}(2020)Hammond, Tsai, \& Pierrehumbert}]{Hammond_2020}
Hammond, M., Tsai, S.-M., \& Pierrehumbert, R.~T. 2020, \apj, 901, 78,
  \dodoi{10.3847/1538-4357/abb08b}

\bibitem[{Haqq-Misra {et~al.}(2018)Haqq-Misra, Wolf, Joshi, Zhang, \&
  Kopparapu}]{Haqq_Misra_2018}
Haqq-Misra, J., Wolf, E.~T., Joshi, M., Zhang, X., \& Kopparapu, R.~K. 2018,
  \apj, 852, 67, \dodoi{10.3847/1538-4357/aa9f1f}

\bibitem[{Held(1983)}]{Held_1983}
Held, I. 1983, In: Hoskins, B.J., Pearce, R. (Eds.), Large-Scale Dynamical
  Processes in the Atmosphere, 127

\bibitem[{Heng \& Showman(2015)}]{Heng_2015}
Heng, K., \& Showman, A.~P. 2015, Annual Review of Earth and Planetary
  Sciences, 43, 509, \dodoi{10.1146/annurev-earth-060614-105146}

\bibitem[{Heng \& Workman(2014)}]{Heng_Workman_2014}
Heng, K., \& Workman, J. 2014, The Astrophysical Journal Supplement Series,
  213, 27, \dodoi{10.1088/0067-0049/213/2/27}

\bibitem[{Herbert {et~al.}(2020)Herbert, Caballero, \& Bouchet}]{Herbert_2020}
Herbert, C., Caballero, R., \& Bouchet, F. 2020, J. Atmos. Sci., 77(1), 31,
  \dodoi{10.1175/JAS-D-19-0089.1}

\bibitem[{Hide(1969)}]{Hide_1969}
Hide, R. 1969, Journal of the Atmospheric Sciences, 26, 841,
  \dodoi{10.1175/1520-0469(1969)026<0841:DOTAOT>2.0.CO;2}

\bibitem[{Holton \& Hakim(2013)}]{Holton_2013}
Holton, J.~R., \& Hakim, G.~J. 2013, An introduction to Dynamic Meteorology:
  5th (London: Academic Press.),
  \dodoi{https://doi.org/10.1016/C2009-0-63394-8}

\bibitem[{Hu \& Yang(2014)}]{Hu_2014}
Hu, Y., \& Yang, J. 2014, Proceedings of the National Academy of Sciences, 111,
  629, \dodoi{10.1073/pnas.1315215111}

\bibitem[{Imamura {et~al.}(2020)Imamura, Mitchell, Lebonnois, Kaspi, \&
  Korablev}]{Imamura_2020}
Imamura, T., Mitchell, J., Lebonnois, S., Kaspi, Y., \& Korablev, O. 2020,
  Space Science Reviews, 216, \dodoi{10.1007/s11214-020-00703-9}

\bibitem[{Kaspi \& Showman(2015)}]{Kaspi_2015}
Kaspi, Y., \& Showman, A.~P. 2015, \apj, 804, 60,
  \dodoi{10.1088/0004-637x/804/1/60}

\bibitem[{Knutson {et~al.}(2007)Knutson, Charbonneau, Allen, Fortney, Agol,
  Cowan, Showman, Cooper, \& Megeath}]{Knutson_2007}
Knutson, H.~A., Charbonneau, D., Allen, L.~E., {et~al.} 2007, Nature, 447, 183,
  \dodoi{10.1038/nature05782}

\bibitem[{Komacek \& Showman(2016)}]{Komacek_2016}
Komacek, T.~D., \& Showman, A.~P. 2016, \apj, 821, 16,
  \dodoi{10.3847/0004-637x/821/1/16}

\bibitem[{Labont{\'{e}} \& Merlis(2020)}]{Labont_2020}
Labont{\'{e}}, M.-P., \& Merlis, T.~M. 2020, \apj, 896, 31,
  \dodoi{10.3847/1538-4357/ab9102}

\bibitem[{Lin {et~al.}(2008)Lin, Mapes, \& Han}]{Lin_2008}
Lin, J.-L., Mapes, B.~E., \& Han, W. 2008, Journal of Climate, 21, 165,
  \dodoi{10.1175/2007JCLI1546.1}

\bibitem[{Lutsko(2018)}]{Lutsko_2018}
Lutsko, N.~J. 2018, Journal of the Atmospheric Sciences, 75, 3,
  \dodoi{10.1175/JAS-D-17-0192.1}

\bibitem[{Matsuno(1966)}]{Matsuno_1966}
Matsuno, T. 1966, J. Meteorol. Soc. Japan, 44, 25,
  \dodoi{10.2151/jmsj1965.44.1_25}

\bibitem[{Merlis \& Schneider(2010)}]{Merlis_2010}
Merlis, T.~M., \& Schneider, T. 2010, Journal of Advances in Modeling Earth
  Systems, 2, \dodoi{10.3894/james.2010.2.13}

\bibitem[{{Noda} {et~al.}(2017){Noda}, {Ishiwatari}, {Nakajima}, {Takahashi},
  {Takehiro}, {Onishi}, {Hashimoto}, {Kuramoto}, \& {Hayashi}}]{Noda_2017}
{Noda}, S., {Ishiwatari}, M., {Nakajima}, K., {et~al.} 2017, Icarus, 282, 1,
  \dodoi{https://doi.org/10.1016/j.icarus.2016.09.004}

\bibitem[{Parmentier \& Crossfield(2018)}]{Parmentier_Crossfield_2018}
Parmentier, V., \& Crossfield, I. J.~M. 2018, In: Deeg H., Belmonte J. (eds)
  Handbook of Exoplanets,
  \dodoi{https://doi.org/10.1007/978-3-319-30648-3_116-1}

\bibitem[{Penn \& Vallis(2017)}]{Penn_2017}
Penn, J., \& Vallis, G.~K. 2017, \apj, 842, 101,
  \dodoi{10.3847/1538-4357/aa756e}

\bibitem[{Perez-Becker \& Showman(2013)}]{Perez-Becker_2013}
Perez-Becker, D., \& Showman, A.~P. 2013, \apj, 776, 134,
  \dodoi{10.1088/0004-637x/776/2/134}

\bibitem[{Phlips \& Gill(1987)}]{Phlips_1987}
Phlips, P.~J., \& Gill, A.~E. 1987, Quarterly Journal of the Royal
  Meteorological Society, 113, 213, \dodoi{10.1002/qj.49711347513}

\bibitem[{Pierrehumbert \& Hammond(2018)}]{Pierrehumbert_2018}
Pierrehumbert, R.~T., \& Hammond, M. 2018, Annu. Rev. Fluid Mech., 51, 275,
  \dodoi{https://doi.org/10.1146/annurev-fluid-010518-040516}

\bibitem[{Shields(2019)}]{Shields_2019}
Shields, A.~L. 2019, The Astrophysical Journal Supplement Series, 243, 30,
  \dodoi{10.3847/1538-4365/ab2fe7}

\bibitem[{Showman {et~al.}(2013)Showman, Fortney, Lewis, \&
  Shabram}]{Showman_2013}
Showman, A.~P., Fortney, J.~J., Lewis, N.~K., \& Shabram, M. 2013, \apj, 762,
  24, \dodoi{10.1088/0004-637x/762/1/24}

\bibitem[{Showman \& Guillot(2002)}]{Showman_2002}
Showman, A.~P., \& Guillot, T. 2002, A\&A, 385, 166,
  \dodoi{10.1051/0004-6361:20020101}

\bibitem[{Showman \& Polvani(2010)}]{Showman_2010}
Showman, A.~P., \& Polvani, L.~M. 2010, Geophys. Res. Lett, 37,
  \dodoi{10.1029/2010GL044343}

\bibitem[{Showman \& Polvani(2011)}]{Showman_2011}
---. 2011, \apj, 738, 71, \dodoi{10.1088/0004-637x/738/1/71}

\bibitem[{{Showman} {et~al.}(2020){Showman}, {Tan}, \&
  {Parmentier}}]{Showman_2020}
{Showman}, A.~P., {Tan}, X., \& {Parmentier}, V. 2020, arXiv e-prints,
  arXiv:2007.15363.
\newblock \doarXiv{2007.15363}

\bibitem[{Stevenson {et~al.}(2014)Stevenson, Desert, Line, Bean, Fortney,
  Showman, Kataria, Kreidberg, McCullough, \& Henry}]{Stevenson_2014}
Stevenson, K.~B., Desert, J.-M., Line, M.~R., {et~al.} 2014, Science, 346, 838,
  \dodoi{10.1126/science.1256758}

\bibitem[{Tsai {et~al.}(2014)Tsai, Dobbs-Dixon, \& Gu}]{Tsai_2014}
Tsai, S.-M., Dobbs-Dixon, I., \& Gu, P.-G. 2014, \apj, 793, 141,
  \dodoi{10.1088/0004-637x/793/2/141}

\bibitem[{Vallis(2006)}]{Vallis_2006}
Vallis, G.~K. 2006, Atmospheric and Oceanic Fluid Dynamics: Fundamentals and
  Large-Scale Circulation (Cambridge, U.K.: Cambridge University Press)

\bibitem[{Van~Tuyl(1986)}]{Tuyl_1986}
Van~Tuyl, A.~H. 1986, Journal of the Atmospheric Sciences, 43,
  \dodoi{10.1175/1520-0469(1986)043<0141:AIOFTM>2.0.CO;2}

\bibitem[{Wu {et~al.}(2001)Wu, Sarachik, \& Battisti}]{Wu_2001}
Wu, Z., Sarachik, E.~S., \& Battisti, D.~S. 2001, Journal of the Atmospheric
  Sciences, 58, 724, \dodoi{10.1175/1520-0469(2001)058<0724:TDTCUR>2.0.CO;2}

\bibitem[{Yang {et~al.}(2019)Yang, Abbot, Koll, Hu, \& Showman}]{Yang_2019}
Yang, J., Abbot, D.~S., Koll, D. D.~B., Hu, Y., \& Showman, A.~P. 2019, \apj,
  871, 29, \dodoi{10.3847/1538-4357/aaf1a8}

\bibitem[{Yang {et~al.}(2013)Yang, Cowan, \& Abbot}]{Yang_2013}
Yang, J., Cowan, N.~B., \& Abbot, D.~S. 2013, \apj, 771,
  \dodoi{10.1088/2041-8205/771/2/l45}

\bibitem[{Zellem {et~al.}(2014)Zellem, Lewis, Knutson, Griffith, Showman,
  Fortney, Cowan, Agol, Burrows, Charbonneau, Deming, Laughlin, \&
  Langton}]{Zellem_2014}
Zellem, R.~T., Lewis, N.~K., Knutson, H.~A., {et~al.} 2014, \apj, 790, 53,
  \dodoi{10.1088/0004-637x/790/1/53}

\bibitem[{Zhang {et~al.}(2018)Zhang, Knutson, Kataria, Schwartz, Cowan,
  Showman, Burrows, Fortney, Todorov, Desert, Agol, \& Deming}]{Zhang_2018}
Zhang, M., Knutson, H.~A., Kataria, T., {et~al.} 2018, \apj, 155, 83,
  \dodoi{10.3847/1538-3881/aaa458}

\bibitem[{Zhang \& Showman(2017)}]{Zhang_2017}
Zhang, X., \& Showman, A.~P. 2017, \apj, 836, 73,
  \dodoi{10.3847/1538-4357/836/1/73}

\end{thebibliography}
\bibliographystyle{aasjournal}








\end{document}